\documentclass[aps,pra,twocolumn,superscriptaddress,notitlepage,showpacs,showkeys]{revtex4-1}
\usepackage{graphicx,amsmath,amsfonts,amssymb,upgreek,txfonts,color}
\usepackage[colorlinks,linkcolor=blue,citecolor=blue,urlcolor=blue,breaklinks=true]{hyperref}
\usepackage{color, colortbl}
\definecolor{lightgreen}{rgb}{0.88,1,1}

\begin{document}



\title{Single-quadrature quantum magnetometry in cavity electromagnonics}

\author{M. S. \surname{Ebrahimi} }
\address{Department of Physics, Faculty of Science, University of Isfahan, Hezar Jerib, 81746-73441, Isfahan, Iran}

\author{Ali \surname{Motazedifard} }
\address{Department of Physics, Faculty of Science, University of Isfahan, Hezar Jerib, 81746-73441, Isfahan, Iran}
\address{Quantum Optics Group, Department of Physics, Faculty of Science, University of Isfahan, Hezar Jerib, 81746-73441, Isfahan, Iran}
\address{Quantum Communication and Quantum Optics group, Iranian Center for Quantum Technologies (ICQTs), Tehran, Iran}

\author{M. Bagheri Harouni}
\email{m.bagheri@sci.ui.ac.ir}
\address{Department of Physics, Faculty of Science, University of Isfahan, Hezar Jerib, 81746-73441, Isfahan, Iran}
\address{Quantum Optics Group, Department of Physics, Faculty of Science, University of Isfahan, Hezar Jerib, 81746-73441, Isfahan, Iran}
\date{\today}

\begin{abstract}
A scheme of an ultrasensitive magnetometer in the cavity quantum electromagnonics is proposed, where the intracavity microwave mode is coupled to a magnonic mode via magnetic dipole interaction. It is shown that by driving both magnonic and microwave modes with external classical fields and controlling the system parameters, one can reduce the added noise of magnetic field measurement below the standard quantum limit (SQL). Surprisingly, we show that beyond the rotating wave approximation (RWA), not only can the added noise be suppressed but also the output cavity response to the input signal can be substantially amplified in order to achieve a precise magnetic field measurement. The estimated theoretical sensitivity of the proposed magnetic amplifier-sensor is approximately on the order of $10^{-18} \rm{T/\sqrt{\rm Hz}}$, which is competitive compared to the current state-of-the-art magnetometers like superconducting quantum interference devices (SQUIDs) and atomic magnetometers. The advantages of the proposed sensor in comparison with the other magnetometers is its high sensitivity at room temperature, sensing in a wide range of frequencies up to MHz, and its capability for signal-response amplification.

\end{abstract}

\maketitle


\section{\label{sec1}Introduction}
The precise measurement of physical quantities like force, magnetic field, etc., is a challenging problem in quantum sensing and metrology since noise sources usually destroy the measurements and cause uncertainty in measurement results. Therefore, applying methods for suppressing or reducing the added noise of measurement, notably below the standard quantum limit (SQL), is an important key in quantum sensing and metrology. As an example, one approach is coherent quantum noise cancellation (CQNC) \cite{1}, in which the antinoise path in the quantum dynamics of the system can cancel the original noise path via destructive quantum interference. Very recently, the backaction noise due to the radiation pressure has been observed in the Advanced Virgo gravitational wave detector on the macroscopic scale at low frequency \cite{virgo2020}.

Recent decades have witnessed theoretical and experimental progress in studying force sensing in different kinds of systems, such as bare and hybrid optomechanical systems \cite{1,2,3,4,5,Review of force sensing}. It has been shown that one can control the quantum noise for simultaneous noise suppression and signal amplification \cite{5}. 
Quantum magnetometry is another important quantum metrology topic with undeniable practical applications in different areas such as geology, navigation, archeology, magnetic storage, medicine \cite{6,7, AVS magnetometry}, and even the search for fundamental physics issues like dark matter axions \cite{one, two, three}. Quantum magnetometry, which has an inevitable role in human life, has been investigated in different kinds of systems such as cavity magnon--polaritons \cite{one,phasemodulatedmagnetometry, 8}, cavity optomechanics \cite{9,10,11,four}, atomic magnetometers \cite{12,13, Noisy atomic magnetometry in real time}, and superconducting quantum interference devices (SQUIDs) \cite{14,15, size of SQUID}. Although subfemto-Tesla magnetic sensing has been achieved in atomic magnetometers and SQUIDs,  there still exist some limitations and challenges such as low-frequency range of sensing \cite{13} and high working temperature \cite{15}.

In recent years, besides the cavity optomechanical systems, studying light-matter interaction based on spin wave collective excitation magnon--photon interaction in ferromagnetic materials like yttrium iron garnet (YIG) with very high spin density and low magnon dissipation rate has attracted a great deal of attention \cite{16,17,18,19,20,21,22,23}. 
Magnons as elementary excitations of magnetically ordered systems can efficiently interact with external magnetic fields and their frequency is tunable, which yields more controllability. These excitations in some ferromagnetic materials like YIG have long lifetimes and long coherence times \cite{24,25,26}, which make them suitable for different applications in quantum information and quantum metrology \cite{16,25}. Very recently, it has been experimentally shown that magnon-based sensing is possible with a high sensitivity through dispersively coupling between the magnetostatic mode and a superconducting qubit \cite{magnon sensing}. A homogeneous magnonic mode can interact with microwave photons via magnetic dipole interaction \cite{16,17,18,Highcooperativity,Nori}. 
It should be noted that YIG has a high Curie temperature as $559$K, and thus is ferromagnetic at both cryogenic and room temperatures \cite{Nori}. Therefore, coupling between a microwave cavity mode and the collective spin excitations is experimentally possible in different YIG samples at either cryogenic \cite{16,18,Highcooperativity,Nori} or room temperature \cite{17,Nori}. 
Microwave photon--magnon interaction that is studied in the context of cavity electromagnonics \cite{28} has been led to some interesting effects including classical Rabi oscillation, magnetically induced transparency, and the Purcell effect \cite{17}. 
Moreover, microwave photon--magnon interaction affects the dynamics of cavity magnomechanics \cite{29, Dynamical Backaction Magnomechanics} where magnons interact with microwave photons and phonons respectively via magnetic dipole and magnetostrictive interactions.
This interaction in different hybrid quantum magnonic-based systems such as cavity electromagnonics and cavity magnomechanics has important effects including microwave-to-optical quantum transducer \cite{30}, quantum entanglement and correlation \cite{31,32,33,34,35,36,last,electromagnonics-optomechanics, foroudcrystalentanglement}, generation of magnon and microwave photon squeezed states \cite{37, microwace field squeezing}, phononic laser \cite{38}, quantum thermometry \cite{39}, quantum magnetometry \cite{one,phasemodulatedmagnetometry,8}, magnon blockade \cite{magnonblockade1,magnonblockade2}, quantum illumination \cite{quantumillumination},  magnon-assisted photon-phonon conversion \cite{photon-phononconversion}, and storage and retrieval of quantum states \cite{Cavity magnomechanical storage and retrieval of quantum
states}.

Motivated by the above-mentioned investigations on force sensing in optomechanics and capability of magnetic sensing in electromagnonic systems as well as the current challenges in magnetometers such as low frequency and high temperature, here, we theoretically propose a feasible experimental scheme of a sensor for high-precision quantum magnetometry. We consider an electromagnonic system whose magnons as quantum excitations of spin wave interact with microwave photons via the magnetic  dipole interaction.
In the proposed system, both magnonic and microwave cavity modes are externally driven by classical fields. We analyze the system both beyond and under the RWA and show that beyond the RWA by controlling system parameters, the added noise of magnetic measurement can be suppressed far below the SQL. In addition, surprisingly, the signal response can be simultaneously amplified, which results in a precise magnetic field measurement. 
The estimated sensitivity of the theoretical proposed magnetometer is on the order of $10^{-18}  \rm{T/\sqrt{\rm Hz}}$, which is competitive or even better than current state-of-the-art SQUIDs and atomic magnetometers. Interestingly,  the advantages of our proposed magnetometer compared to the SQUIDs and atomic based magnetometers are that our magnetometer can even operate at \textit{room temperature} and over a \textit{wide range} of frequency with high sensitivity and precision.

The paper is organized as follows. In Sec. \ref{sec2}, the system Hamiltonian is described, and then in Sec. \ref{sec3},  the dynamics of the system is investigated through the quantum Langevin-Heisenberg equations of motion. The mechanisms of the magnetic sensing, noise suppression, and signal amplification are presented in Sec. \ref{sec4}. In Sec. \ref{sec5}, the sensitivity, the signal-to-noise ratio (SNR), and comparison to other magnetometers are discussed. Finally, the summary, conclusion, and outlooks are mentioned in Sec. \ref{sec6}.

\section{\label{sec2} The system and Hamiltonian}

\begin{figure}
\includegraphics[width=8cm]{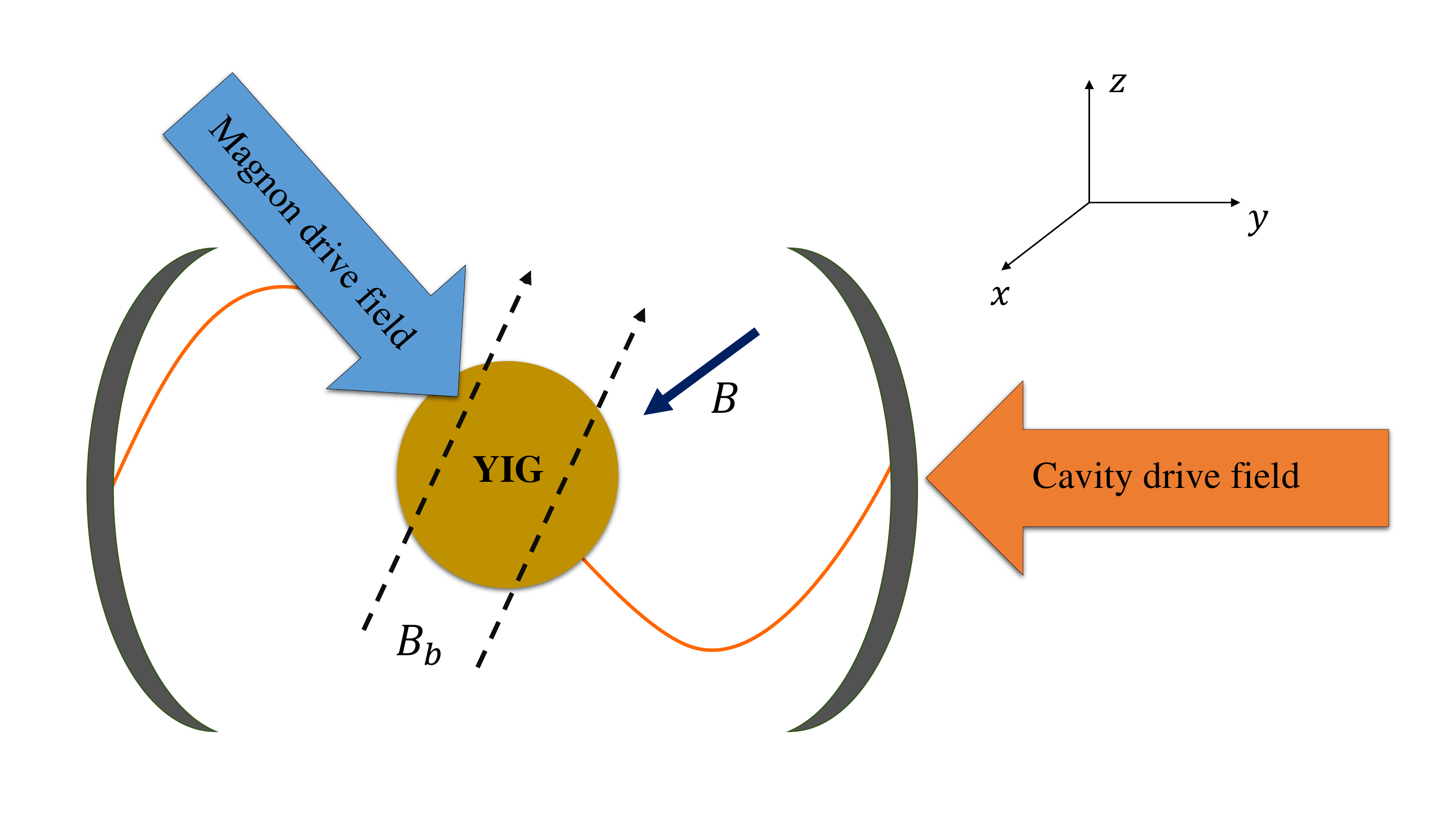}
\caption{(Color online) Schematic of a magnetic field sensor based on cavity electromagnonics, in which a YIG sphere is placed where the magnetic field of the microwave cavity mode is maximum. The YIG sphere is also biased with a uniform magnetic field $B_b$ in a direction which at least has a component along the direction orthogonal to the cavity axis. The cavity and magnonic modes are driven by the classical cavity and magnon drive fields, respectively. $B$ is an unknown weak external magnetic field which is applied to the YIG sphere along the $x$ direction and should be measured. }
\label{Fig1}
\end{figure}

Figure \ref{Fig1}  shows a schematic of the proposed magnetometer: A YIG sphere is placed inside a microwave cavity with a uniform external bias magnetic field, $B_b$, which is applied to the YIG sphere to produce a homogeneous magnonic mode, the so-called Kittel mode. Therefore, magnetic dipole interaction can mediate microwave photon--magnon interaction \cite{16,17,27,28}. Assuming that both the magnonic and photonic modes are driven by external magnetic and electromagnetic fields, respectively, the Hamiltonian of the system  is given by
\begin{eqnarray}\label{1}
\hat{H}&=& \hbar \omega_a \hat{a}^{\dagger}\hat{a}+\hbar \omega_m \hat{m}^{\dagger}\hat{m}+\hbar g_0 (\hat{a}+\hat{a}^{\dagger})(\hat{m}+\hat{m}^{\dagger})\nonumber\\
&&+i\hbar \epsilon_L( \hat{a}^{\dagger} e^{-i\omega_L t}-\hat{a}e^{i\omega_L t})+i\hbar \epsilon_d( \hat{m}^{\dagger} e^{-i\omega_d t}-\hat{m}e^{i\omega_d t})\nonumber\\
&&+\hat{H}_B.\
\end{eqnarray}

The first two terms describe the free energy of photonic and magnonic modes where $\hat{a}$ $(\hat{a}^{\dagger})$ and $\hat{m}$ $(\hat{m}^{\dagger})$ respectively denote the annihilation (creation) operators of these two modes and satisfy bosonic algebra $[\hat{o},\hat{o}^{\dagger}]=1$, $\hat{o}=\hat{a},\hat{m} $.  $\omega_a$ and $\omega_m$ indicate the resonance frequency of the cavity and magnonic modes, respectively. It is worth noting that the magnon frequency is determined by the external bias magnetic field $B_b$ and the gyromagnetic ratio $\gamma/2\pi = {28} ~\mathrm{GHz/ T}$, i.e., $\omega_m=\gamma B_b$, and can be easily adjusted via $ B_b $.
The third term in the Hamiltonian represents the microwave photon--magnon coupling obtained through the magnetic dipole interaction, which is written without resorting to the rotating wave approximation (RWA) \cite{27,28}. Note that the RWA breaks in the ultrastrong coupling (USC) and deep strong coupling (DSC) regimes, which respectively can be achieved when conditions $g_0 \lesssim \omega_{a(m)}$ and $ g_0 \gtrsim  \omega_{a(m)}$ are satisfied \cite{USC}. 
Therefore, this interaction Hamiltonian can be considered for either USC or DSC regime. Note that the strength of magnon-photon coupling is given by $g_0=\frac{\gamma B_0}{2}\sqrt{2sN}$, where $\gamma$ is the gyromagnetic ratio and $B_0=\sqrt{\frac{\hbar \omega_a \mu_0}{V_a}}$ indicates the amplitude of the magnetic field inside the microwave cavity in which  $\omega_a$ is the microwave mode frequency,  $\mu_0$ is the vacuum permeability, and $V_a$ stands for the mode volume of the microwave cavity resonance.  Also, $s$ and $N$, respectively, stand for the spin angular momentum $s=5/2$ on each unit cell of the magnetic material and the number of unit cells with spin $s$ in the YIG sphere \cite{27}.
The contribution of the two external drive fields can be seen in the fourth and fifth terms of the Hamiltonian. The cavity drive field with frequency $\omega_L$ and power $P_{in}$ drives the cavity mode with rate $\epsilon_L=\sqrt{\frac{2P_{in}\kappa_a}{\hbar\omega_L}}$ (with $\kappa_a$ being the cavity decay rate). The coupling strength of the magnon drive field (whose frequency and amplitude are respectively given by $\omega_d$ and $B_d$) with the magnonic mode in the YIG sphere is  $\epsilon_d = \frac{\eta}{2} B_d$, where $\eta = \frac{\gamma}{2} \sqrt{5N}$ \cite{31}. 
 $\hat{H}_B$ accounts for the coupling of the magnonic mode in the YIG sphere to an unknown weak external magnetic field $B$, which should be measured, and is given by
\begin{equation}\label{2}
\hat{H}_B=-\gamma \vec{S}\cdot \vec{B}=-\hbar \eta B(t) (\hat{m}+\hat{m}^{\dagger}),
\end{equation}
with $\vec{S}$ being the collective spin angular momentum. The above Hamiltonian is obtained using the Holstein-Primakkoff transformations and  assuming that the time-dependent external magnetic field is along the $x$ direction~\cite{31}.
The Hamiltonian in the frame rotating at frequencies $\omega_L$ and $ \omega_d $ is given by
\begin{eqnarray}\label{4}
\hat{H}^{(rot)}&=&\hbar\Delta_a\hat{a}^{\dagger}\hat{a}+\hbar\Delta_m\hat{m}^{\dagger}\hat{m}+\hbar g_0 (\hat{a}e^{-i\omega_L t}+\hat{a}^{\dagger}e^{i\omega_L t})\nonumber\\
&&(\hat{m}e^{-i\omega_d t}+\hat{m}^{\dagger}e^{i\omega_d t})+i\hbar\epsilon_L(\hat{a}^{\dagger}-\hat{a})+i\hbar \epsilon_d(\hat{m}^{\dagger}-\hat{m})\nonumber\\
&&-\hbar \eta B(t) (\hat{m}e^{-i\omega_d t}+\hat{m}^{\dagger}e^{i\omega_d t}),\
\end{eqnarray}
where $\Delta_a \equiv \omega_a-\omega_L$ and $\Delta_m\equiv \omega_m-\omega_d$ are the cavity and magnonic detunings, respectively.
We assume that the coupling strength $ g_0 $ is modulated coherently as  $g_0(t)= g (1+2\cos \Omega t)$ with $ g=g_0\mathcal{E} $ where $\mathcal{E}$ and $\Omega$ are the amplitude and the frequency of time modulation, respectively. 
Assuming resonance condition, $\Omega= \omega_L+\omega_d$, and also considering the fact that the magnonic and microwave modes are near resonance in the microwave photon--magnon interaction $ (\omega_a\simeq\omega_m )$ \cite{17} and consequently  $ \omega_L\simeq\omega_d $ can be provided, then, $\Omega=2\omega_L$. In this manner, the photon--magnon interaction term [the third term in the Hamiltonian in Eq. (\ref{4})], by considering the time modulation $ g_0(t) $ and using the RWA, which is valid when $g \equiv g_0\mathcal{E} \ll \Omega$, can be simplified as 
\begin{equation}\label{5}
\hat{H}_{int}=\hbar g (\hat{a}+\hat{a}^{\dagger})(\hat{m}+\hat{m}^{\dagger}).
\end{equation}
As mentioned, the magnon--photon coupling strength $g_0$ depends on the number of spins and the amplitude of the magnetic field inside the microwave cavity. Therefore, by modulating the amplitude of magnetic field inside the microwave cavity, one can modulate the magnon--photon coupling strength. By controlling the modulation amplitude of magnetic field, we can change $\mathcal{E}$ and hence $g$ so that $g \ll \ \Omega$ is valid.
 On the other hand, the effective magnon--photon coupling strength can be tuned via the modulation amplitude $ \mathcal{E} $. This controllability on coupling strength is an advantage of the time modulation in this configuration. It should be noted that recently the time modulation technique has been theoretically and experimentally applied in quantum optical systems in order to generate the nonclassical state, squeezing and also improving the sensing \cite{3,4,5,aliDCEsqueezing,aliDCE1,aliDCE2,aliDCE3,NoriDCE1,NoriDCE2,NoriDCE3,pontinmodulation}.

Now, the total Hamiltonian of the system is given by
\begin{eqnarray}\label{6}
\hat{H}&=&\hbar\Delta(\hat{a}^{\dagger}\hat{a}+\hat{m}^{\dagger}\hat{m})+\hbar g (\hat{a}+\hat{a}^{\dagger})(\hat{m}+\hat{m}^{\dagger})\nonumber\\
&&+i\hbar\epsilon_L(\hat{a}^{\dagger}-\hat{a})+i\hbar \epsilon_d(\hat{m}^{\dagger}-\hat{m})-\hbar \eta B(t) (\hat{m}e^{-i\omega_d t}+\hat{m}^{\dagger}e^{i\omega_d t}),\nonumber\\
\end{eqnarray}
in which we have assumed $\Delta_a\simeq\Delta_m\equiv \Delta$ since $\omega_a \simeq \omega_m$ and $\omega_L \simeq \omega_d$ and we have considered the first notation of Hamiltonian ($\hat{H}$ instead of $\hat{H}^{(rot)}$).

\section{\label{sec3} DYNAMICS OF THE SYSTEM}
Considering Hamiltonian (\ref{6}), the quantum Langevin equations (QLEs) describing the system dynamics are given by
\begin{eqnarray}
\delta \dot{\hat{m}}&=&-i\Delta \delta \hat{m}-ig(\delta \hat{a}^{\dagger}+\delta \hat{a})-\frac{\kappa_m}{2}\delta\hat{m}+\sqrt{\kappa_m}\hat{m}^{\mathrm{in}}\nonumber\\
&&+i\eta B(t) e^{i\omega_d t}, \label{7}\\
\delta \dot{\hat{a}}&=&-i\Delta \delta \hat{a}-ig(\delta\hat{m}^{\dagger}+\delta \hat{m})-\frac{\kappa_a}{2}\delta\hat{a}+\sqrt{\kappa_a}\hat{a}^{\mathrm{in}}, \label{8}
\end{eqnarray}
in which $\kappa_a$ and $\kappa_m$ indicate the cavity and magnonic dissipation rates, respectively. The input quantum noise of the cavity and magnonic modes are respectively shown by $\hat{a}^{\mathrm{in}}$ and $\hat{m}^{\mathrm{in}}$ which satisfy the following correlation functions
\begin{eqnarray}
\langle \hat{o}^{\mathrm{in}}(t)\hat{o}^{\mathrm{in} \dagger}(t^\prime)\rangle &=&(1+\bar{n}_o)\delta(t-t^{\prime}), \label{9} \\
\langle \hat{o }^{\mathrm{in} \dagger}(t)\hat{o}^{\mathrm{in} }(t^\prime)\rangle &=&\bar{n}_o\delta(t-t^{\prime}), \label{10}
\end{eqnarray}
with $o=a,m$ and $\bar{n}_o=[\rm exp(\frac{\hbar \omega_o}{k_B T})-1]^{-1}$ being the mean number of thermal excitations of the cavity and magnon modes at temperature $T$. 
Here, we assume $\bar{n}_a\simeq\bar{n}_m\equiv\bar{n}$, which is experimentally realizable for the microwave photon--magnon interaction $\omega_a\simeq\omega_m$ \cite{17}. By defining the quadratures $\delta \hat{X}_o=\frac{\hat{o}+\hat{o}^{\dagger}}{\sqrt{2}}$ and $\delta \hat{P}_o=\frac{i(\hat{o}^{\dagger}-\hat{o})}{\sqrt{2}}$ $(\hat{o}=\delta \hat{a}, \delta \hat{m})$, Eqs.~(\ref{7}) and (\ref{8}) can be written as follows:
\begin{eqnarray}
 \delta \dot{\hat{X}}_m&=&\Delta \delta\hat{P}_m-\frac{\kappa_m}{2}\delta \hat{X}_m+\sqrt{\kappa_m}\hat{X}^{\prime \mathrm{in}}_m, \label{11} \\
\delta \dot{\hat{P}}_m&=&-\Delta \delta \hat{X}_m-2g\delta \hat{X}_a-\frac{\kappa_m}{2}\delta \hat{P}_m+\sqrt{\kappa_m}\hat{P}^{\prime \mathrm{in}}_m, \label{12} \\
\delta \dot{\hat{X}}_a&=&\Delta \delta \hat{P}_a-\frac{\kappa_a}{2}\delta \hat{X}_a+\sqrt{\kappa_a}\hat{X}^{\mathrm{in}}_a, \label{13} \\
\delta \dot{\hat{P}}_a&=&-\Delta \delta\hat{X}_a-2g\delta \hat{X}_m-\frac{\kappa_a}{2}\delta \hat{P}_a+\sqrt{\kappa_a}\hat{P}^{\mathrm{in}}_a,  \label{14} 
\end{eqnarray}
where $\hat{X}^{\mathrm{in}}_o=\frac{\hat{o}^{\mathrm{in}}+\hat{o}^{\mathrm{in} \dagger}}{\sqrt{2}}$ and $\hat{P}^{\mathrm{in}}_o=\frac{i(\hat{o}^{\mathrm{in} \dagger}-\hat{o}^{\mathrm{in}})}{\sqrt{2}}$ ($o= a, m$). Also, we have defined $\hat{X}^{\prime \mathrm{in}}_m(t)$ and $\hat{P}^{\prime \mathrm{in}}_m(t)$ as follows:
\begin{eqnarray}
\hat{X}^{\prime \mathrm{in}}_m(t)&=&\hat{X}^{\mathrm{in}}_m-\sqrt{\frac{2}{\kappa_m}}\eta B(t) \sin (\omega_d t), \label{15}\\
\hat{P}^{\prime \mathrm{in}}_m(t)&=&\hat{P}^{\mathrm{in}}_m+\sqrt{\frac{2}{\kappa_m}}\eta B(t) \cos(\omega_d t), \label{16}
\end{eqnarray}
which show the effects of the external magnetic field.

In the system under consideration, the external magnetic field that we are going to measure, $B$, is coupled to the magnonic mode. On the other hand, the magnon mode is coupled to microwave photons through the magnetic dipole interaction. Consequently, the signal corresponding to the external magnetic field, $B$, can be detected by measuring the optical output phase quadrature, $\hat{P}^{\mathrm{out}}_a$. According to the input-output relation, i.e., $\hat{a}^{\mathrm{out}}=\sqrt{\kappa_a}\delta \hat{a}-\hat{a}^{\mathrm{in}}$, the output phase quadrature is given by 
\begin{equation}\label{17}
\hat{P}^{out}_a=\sqrt{\kappa_a}\delta \hat{P}_a-\hat{P}^{in}_a.
\end{equation}

By solving the QLEs [Eqs. (\ref{11}) -(\ref{14})] in the Fourier space, we can find $\delta \hat{P}_a(\omega)$. Then, according to Eq.(\ref{17}), $\hat{P}^{\mathrm{out}}_a$ in the Fourier space can be simplified as
\begin{eqnarray}\label{18}
\hat{P}^{\mathrm{out}}_a(\omega)&=& \mathcal{A}(\omega)\hat{X}^{\prime \mathrm{in}}_m(\omega)+\mathcal{B}(\omega)\hat{P}^{\prime \mathrm{in}}_m(\omega)\nonumber\\
&&+\mathcal{C}(\omega)\hat{X}^{\mathrm{in}}_a(\omega)+\mathcal{D}(\omega)\hat{P}^{\mathrm{in}}_a(\omega),
\end{eqnarray}
where $\hat{X}^{\prime \mathrm{in}}_m(\omega)$ and $\hat{P}^{\prime \mathrm{in}}_m(\omega)$ are respectively the Fourier transform of $\hat{X}^{\prime \mathrm{in}}_m(t)$ and $\hat{P}^{\prime \mathrm{in}}_m(t)$ and are given by
\begin{eqnarray}
\hat{X}^{\prime \mathrm{in}}_m(\omega)&=&\hat{X}^{\mathrm{in}}_m (\omega)+\sqrt{\frac{2}{\kappa_m}}\eta \frac{i}{2} [B(\omega+\omega_d)-B(\omega-\omega_d)], \label{19} \\
\hat{P}^{\prime \mathrm{in}}_m(\omega)&=&\hat{P}^{\mathrm{in}}_m (\omega)+\sqrt{\frac{2}{\kappa_m}}\eta \frac{1}{2} [B(\omega+\omega_d)+B(\omega-\omega_d)]. \label{20}
\end{eqnarray}

In addition, the coefficients $\mathcal{A}(\omega), \mathcal{B}(\omega), \mathcal{C}(\omega)$, and $\mathcal{D}(\omega)$ are defined as
\begin{eqnarray}
\mathcal{A}(\omega)&\equiv & 4g\sqrt{\kappa_a \kappa_m}(\frac{\Delta\chi_a(\omega)-1}{2i\omega+\kappa_a})\chi^{\prime}_m(\omega), \label{21}\\
\mathcal{B}(\omega)&\equiv & 4g\Delta\chi_m(\omega)\chi^{\prime}_m(\omega)\sqrt{\kappa_a\kappa_m}(\frac{\Delta\chi_a(\omega)-1}{2i\omega+\kappa_a}), \label{22}\\
\mathcal{C}(\omega)&\equiv & -4g^2\Delta\chi_m(\omega)\chi^{\prime}_m(\omega)\chi_a(\omega)\kappa_a (\frac{\Delta\chi_a(\omega)-1}{\Delta})\nonumber\\
&&-\kappa_a\chi_a(\omega), \label{23}\\
\mathcal{D}(\omega)&\equiv & -8g^2\Delta\chi_m(\omega)\chi^{\prime}_m(\omega)\chi_a(\omega)\kappa_a (\frac{\Delta\chi_a(\omega)-1}{2i\omega+\kappa_a})\nonumber\\
&&-\frac{2\kappa_a(\Delta\chi_a(\omega)-1)}{2i\omega+\kappa_a}-1, \label{24}
\end{eqnarray}
where we have defined $\chi_a(\omega)$, $\chi_m(\omega)$, and $\chi^{\prime}_m(\omega)$ as follows:
\begin{eqnarray}
\chi_a(\omega) &\equiv &\frac{\Delta}{\Delta^2-\omega^2+i\omega\Delta +\frac{\kappa_a^2}{4}}, \label{25}\\
\chi_m(\omega)&\equiv &\frac{1}{i\omega+\frac{\kappa_m}{2}}, \label{26}\\
\chi^{\prime}_m(\omega)&\equiv &\chi_m(\omega)[1+\Delta^2\chi^2_m(\omega)-4\Delta g^2\chi_a(\omega)\chi^2_m(\omega)]^{-1}.\label{27}
\end{eqnarray}

\section{\label{sec4} MAGNETIC FIELD SENSING}
Generally, similar to the cavity quantum electrodynamics systems, there could be different kinds of light-matter interaction in cavity electromagnonic systems \cite{ultra strong coupling regime nori,17} such as DSC, USC, strong coupling (SC), and weak coupling (WC) regimes, which depend on the light-matter coupling strength, $g_0$. 
The DSC and USC regimes are satisfied when $ g_0 \gtrsim \omega_{a(m)}$ and $g_0 \lesssim \omega_{a(m)}$, respectively. Note that the RWA breaks down in these two regimes. 
In SC and WC regimes, in addition to the frequency and coupling, the damping rates play the important role such that they can be labeled by $\kappa_a, \kappa_m\ \ll\ g_0 \ll\ \omega_{a(m)}$ and $g_0 \ll\ \kappa_a, \kappa_m, \omega_{a(m)}$, respectively. It should be noted that in these two later regimes the RWA is valid. 
Furthermore, the intermediate coupling (IMC) regime where also the RWA is valid is satisfied when  $(g_0\ll\ \omega_{a(m)}$, $\kappa_a < \ g_0 < \ \kappa_m)$ or $(g_0 \ll\ \omega_{a(m)}$, $\kappa_m < \ g_0 < \ \kappa_a)$.

In the following, we first investigate the magnetic field sensing in the cavity electromagnonics beyond the RWA, which can be considered for the DSC or USC regime. Then, we study the possibility of magnetic sensing under the RWA, which is valid in SC, WC, and IMC regimes.\\

\subsection{\label{secA} Beyond the RWA}

In this section, we consider the system beyond the RWA and show  that the signal response  corresponding to the external magnetic field $B$ can be detected via the output cavity phase-quadrature $P^{\mathrm{out}}_a$. Therefore, we calculate the output cavity phase spectrum and show how by selecting suitable values of system parameters, one can simultaneously suppress the added noise and amplify the response to the input signal.

The spectrum of the optical output phase quadrature that helps to detect the input magnetic field is as follows:
\begin{eqnarray}\label{28}
S_{P^{\mathrm{out}}_a}(\omega)&=&\frac{1}{4\pi}\int d\omega^{\prime}e^{i(\omega+\omega^{\prime})t}\langle \delta P^{\mathrm{out}}_a(\omega)\delta P^{\mathrm{out}}_a(\omega^{\prime})\nonumber\\
&&+\delta P^{\mathrm{out}}_a(\omega^{\prime})\delta P^{\mathrm{out}}_a(\omega)\rangle.
\end{eqnarray}

Let us now set aside the external magnetic field, $B$, for simplification since the response of the system to the external magnetic field is independent of the classical input magnetic field and only depends on the quantum properties of the system. In this manner, the output phase quadrature spectrum is obtained as
\begin{eqnarray}\label{29}
S_{P^{\mathrm{out}}_a}(\omega)&=&[\vert \mathcal{A}(\omega)\vert ^2+\vert \mathcal{B}(\omega)\vert ^2](\bar{n}_m+\frac{1}{2})\nonumber\\
&&+ [\vert \mathcal{C}(\omega)\vert ^2+\vert \mathcal{D}(\omega)\vert ^2](\bar{n}_a+\frac{1}{2}).
\end{eqnarray}
We can rewrite this spectrum as follows
\begin{equation}\label{30}
S_{P^{\mathrm{out}}_a}(\omega)=R_m(\omega)[(\bar{n}_m+\frac{1}{2})+n_{\mathrm{add}}(\omega)],
\end{equation}
in which $R_m(\omega)$ and $n_{\mathrm{add}}(\omega)$ show respectively the magnonic response and the added noise of magnetic field measurement beyond the RWA and are given by
\begin{eqnarray}
R_m(\omega)&=&[\vert \mathcal{A}(\omega)\vert ^2+\vert \mathcal{B}(\omega)\vert ^2], \label{31}\\
n_{\mathrm{\mathrm{add}}}(\omega)&=&(\bar{n}_a+\frac{1}{2})\frac{ [\vert \mathcal{C}(\omega)\vert ^2+\vert \mathcal{D}(\omega)\vert ^2]}{[\vert \mathcal{A}(\omega)\vert ^2+\vert \mathcal{B}(\omega)\vert ^2]}. \label{32}
\end{eqnarray}
So, according to Eqs. (\ref{21})--(\ref{24}), the magnonic response and the added noise can be simplified as
\begin{widetext}
\begin{eqnarray}
R_m(\omega) & =& 16 g^2\kappa_a\kappa_m\vert \chi^{\prime}_m(\omega)\vert ^2 \big \vert \frac{\Delta \chi_a(\omega)-1}{2i\omega+\kappa_a} \big \vert ^2 \big[\Delta^2\vert \chi_m(\omega)\vert^2 \! + \! \! 1 \big], \label{33}\\
n_{\mathrm{add}}(\omega) & = & \frac{\Big(\big \vert 8g^2\kappa_a \Delta\chi_m(\omega)\chi^{\prime}_m(\omega)\chi_a(\omega)(\frac{\Delta\chi_a(\omega)-1}{2i\omega+\kappa_a})
 +1+\frac{2\kappa_a(\Delta\chi_a(\omega)-1)}{2i\omega+\kappa_a}\big \vert^2+\big \vert 4g^2\kappa_a \chi_m(\omega)\chi^{\prime}_m(\omega)\chi_a(\omega)
 (\Delta \chi_a(\omega)-1)+\kappa_a\chi_a(\omega)\big \vert ^2\Big)(\bar{n}_a+\frac{1}{2})}
 {16g^2\kappa_a\kappa_m \vert \chi^{\prime}_m(\omega)\vert^2\big \vert (\frac{\Delta \chi_a(\omega)-1}{2i\omega+\kappa_a})\big \vert ^2(\Delta^2\vert \chi_m(\omega)\vert ^2+1)}. \nonumber\\\label{34} 
\end{eqnarray}
\end{widetext}

\begin{figure}
\includegraphics[width=8cm]{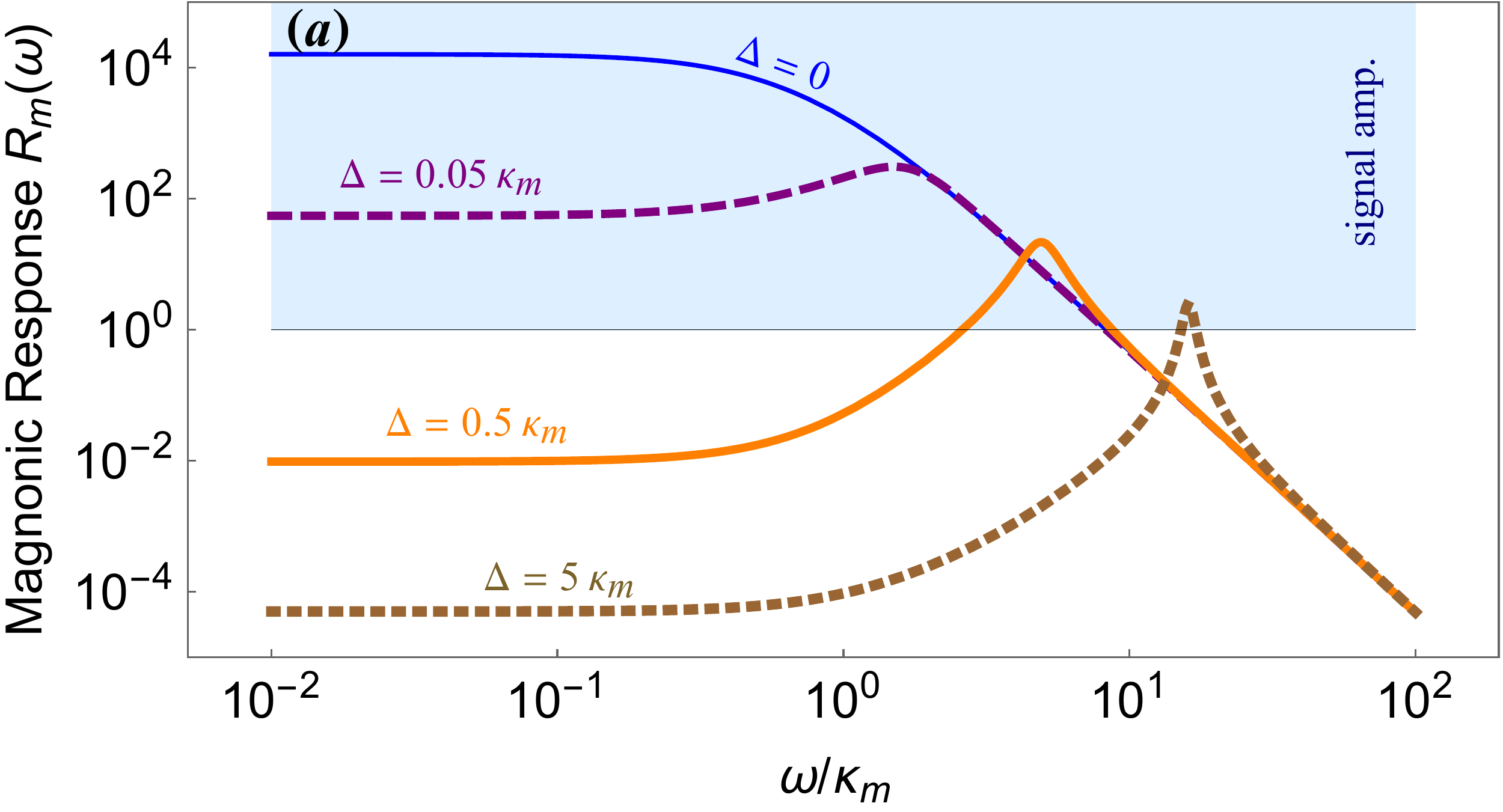}
\hspace{10mm}
\includegraphics[width=8cm]{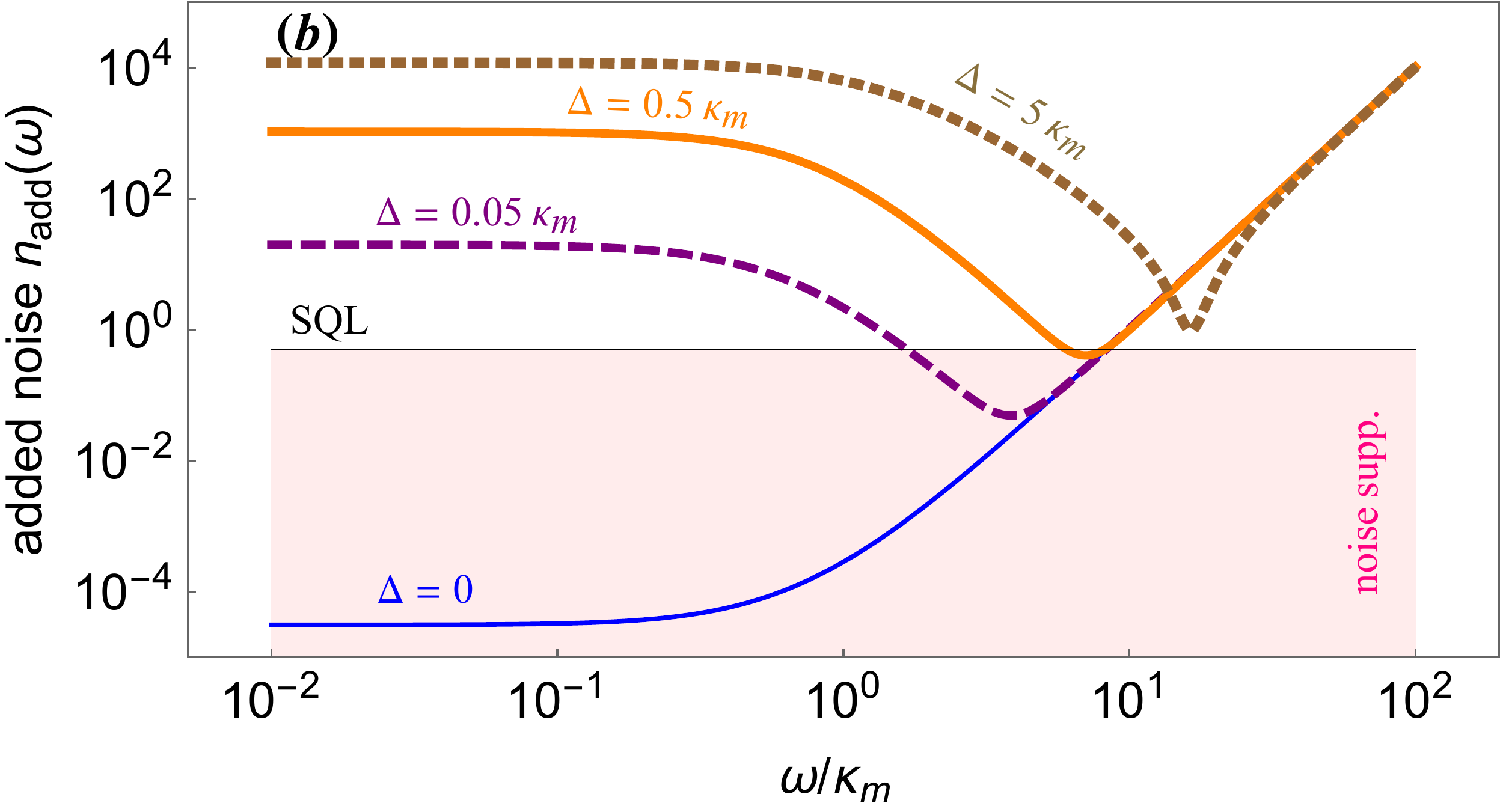}
\caption{(Color online) (a) The magnonic response to the input signal, $R_m(\omega)$, and (b) the added noise of magnetic field measurement, $n_{\mathrm{add}}(\omega)$, vs the normalized frequency $\omega/\kappa_m$. All curves in these two panels have been plotted beyond the RWA corresponding to the cooperativity $C=1000$. The thin blue solid, purple dashed, thick orange solid, and brown dotted curves, respectively, correspond to detunings $\Delta=0,0.05\kappa_m, 0.5\kappa_m, 5\kappa_m$. Here, we have set zero thermal noises, $\bar{n}\simeq 0$, corresponding to the zero temperature. }
\label{Fig2}
\end{figure}

\begin{figure}
\includegraphics[width=8cm]{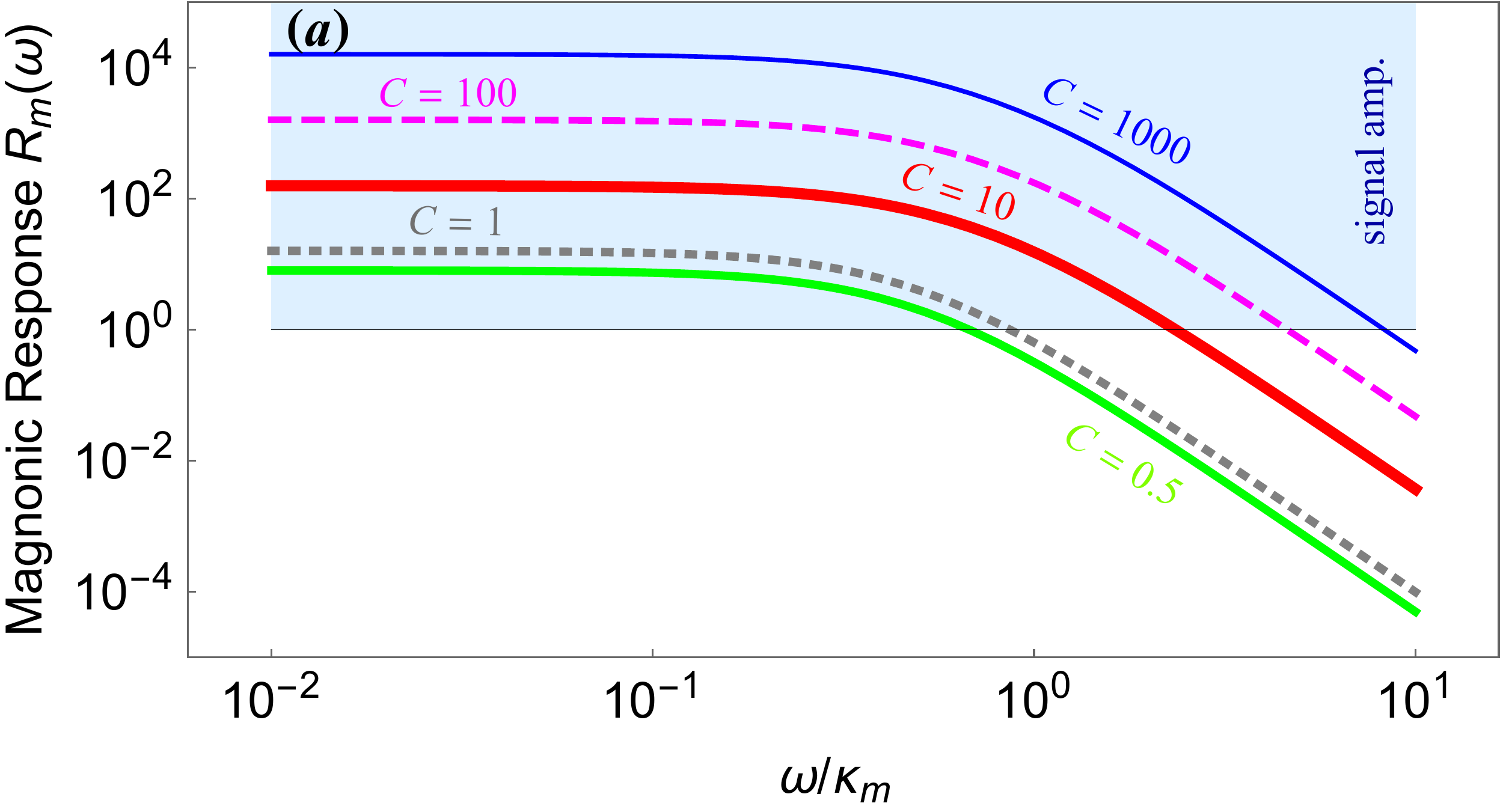}
\hspace{10mm}
\includegraphics[width=8cm]{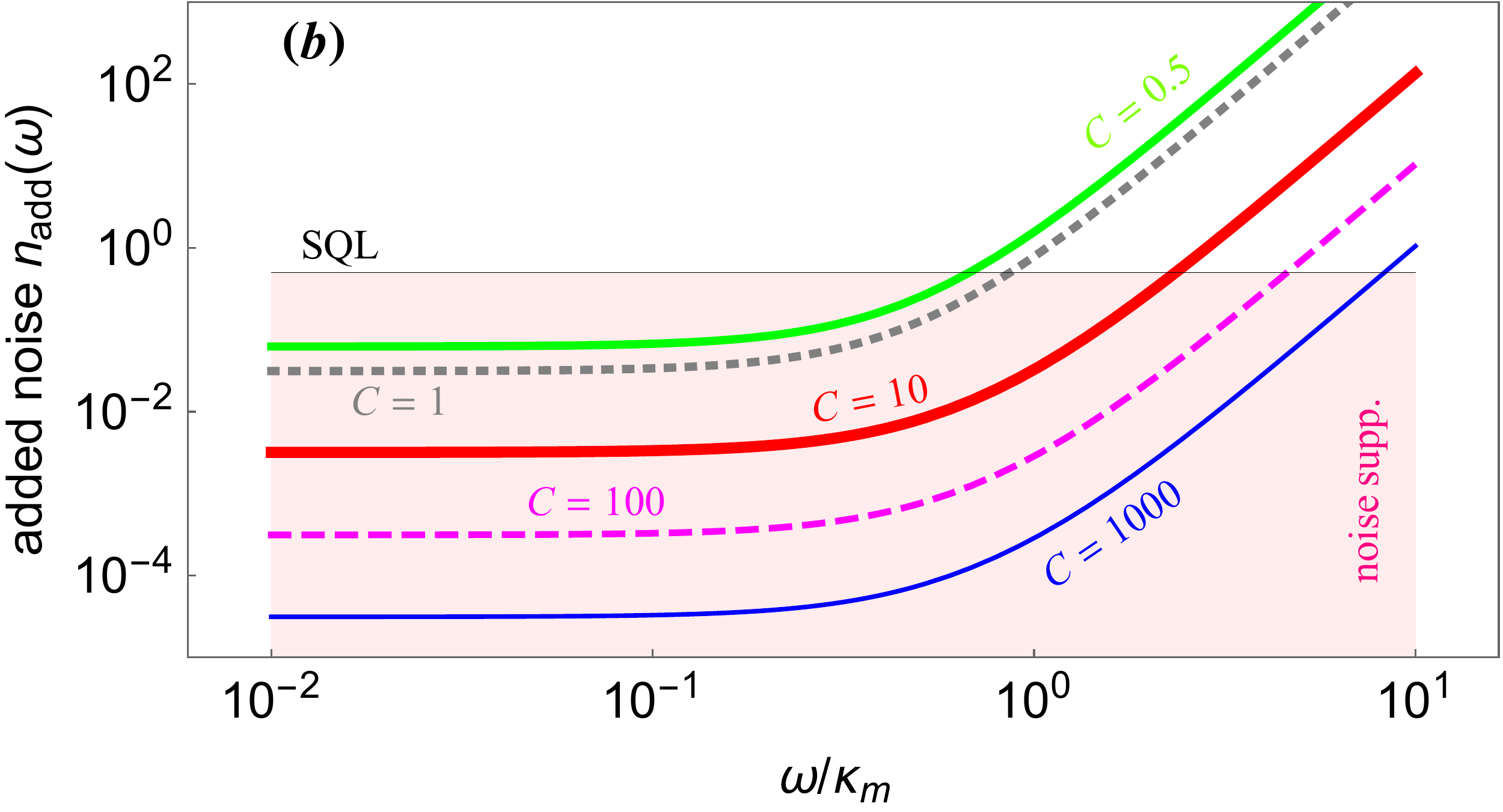}
\caption{(Color online) (a) The magnonic response to the input signal, $R_m(\omega)$, and (b) the added noise of magnetic field measurement, $n_{\mathrm{add}}(\omega)$, versus the normalized frequency $\omega/\kappa_m$. All curves have been plotted  beyond the RWA  for the \textit{optimized} zero detuning, $\Delta=0$. The thin blue solid, magenta dashed, thick red solid, gray dotted, and green solid curves are respectively referred to $C=1000, 100, 10, 1, 0.5$. Here, we have set zero thermal noises, $\bar{n}\simeq 0$, corresponding to the zero temperature. }
\label{Fig3}
\end{figure}

First, let us consider the on-resonance case $\omega=0$. By considering  Eqs.~(\ref{33}) and (\ref{34}), one finds the \textit{optimized} condition $\Delta=0$ in order to be able simultaneously amplify the signal and suppress the noise. In this manner, the magnonic response and the added noise are given by
\begin{eqnarray}
R_m(\omega=0)\vert_{\Delta=0}&=&\frac{64g^2}{\kappa_a\kappa_m}=16C, \label{35}\\
n_{\mathrm{add}}(\omega=0)\vert_{\Delta=0}&=&(\bar{n}_a+\frac{1}{2})\frac{\kappa_a\kappa_m}{64g^2}=\frac{(\bar{n}_a+\frac{1}{2})}{16C}, \label{36}
\end{eqnarray}
in which we have defined $C\equiv\frac{4g^2}{\kappa_a\kappa_m}$ as the electromagnonic cooperativity beyond the RWA.
It is evident that in the on-resonance case, one can linearly amplify the response and suppress the noise, simultaneously, by increasing the electromagnonic cooperativity $C$. Note that beyond the RWA, the cooperativity is controlled via the effective coupling $ g=g_0 \mathcal{E} $ which itself can be controlled via the modulation amplitude of the external magnetic drive.

In order to obtain more insights about the effect of different values of detuning, $\Delta$, on signal amplification and noise suppression beyond the RWA, in Fig. \ref{Fig2}  the magnonic response of the cavity output phase spectrum $R_m(\omega)$ [Fig. \ref{Fig2}(a)] and the added noise of magnetic field measurement $n_{\mathrm{add}}(\omega)$ [Fig. \ref{Fig2}(b)] have been plotted versus the normalized frequency $\omega/\kappa_m$ for  different values of detuning. As is evident, at zero temperature, $\bar{n}=0$, and for an arbitrary selected electromagnonic cooperativity, for example $C=1000$, by choosing the frequency of the external drive fields equal to the cavity field frequency, the magnonic response can be increased to high value for a wide range of frequency (for $\omega \lesssim \kappa_m$) and simultaneously the added noise of magnetic field measurement can be suppressed below the SQL. As the results show, by increasing the detuning the amount of signal amplification and noise reduction will decrease. For small detunings like $\Delta =0.05\kappa_m$, the frequency range in which we could at the same time amplify the signal and suppress the added noise is very short. While for larger detunings like $\Delta=5\kappa_m$, both signal amplification and noise reduction are impossible. Therefore, the \textit{optimized} case in which we could amplify the signal and simultaneously suppress the added noise of measurement, especially for a wide range of frequency is zero detuning $\Delta=0$.
As is evident, although simultaneous signal amplification and noise reduction have the maximum value in the on-resonance frequency,  they are still possible and appreciable far from the on-resonance case.

In Fig. \ref{Fig3}, we have plotted the magnonic response of the cavity output phase spectrum $R_m(\omega)$ [Fig. \ref{Fig3}(a)] and the added noise of magnetic field measurement $n_{add}(\omega)$ [Fig. \ref{Fig3}(b)] versus the normalized frequency $\omega/ \kappa_m$ for different cooperativities in order to see how different values of cooperativity affect the signal amplification and noise reduction. These results that are obtained for the \textit{optimized} detuning, $\Delta=0$, and at zero temperature show that by increasing the electromagnonic cooperativity signal amplification and noise reduction, which occur simultaneously, will increase. 
Although the large cooperativity regime like $ C=1000 $ has more signal amplification and noise suppression,  the weak cooperativity regime such as $C=0.5$ can also simultaneously lead to signal amplification and noise suppression in both on-resonance and off-resonance frequencies.
Note that beyond the RWA, using the coherent time modulation of amplitude of magnetic drive which yields to the modulation of the magnon--photon coupling, one can control and tune the electromagnonic cooperativities via the amplitude of modulation.



\begin{figure}
\includegraphics[width=8cm]{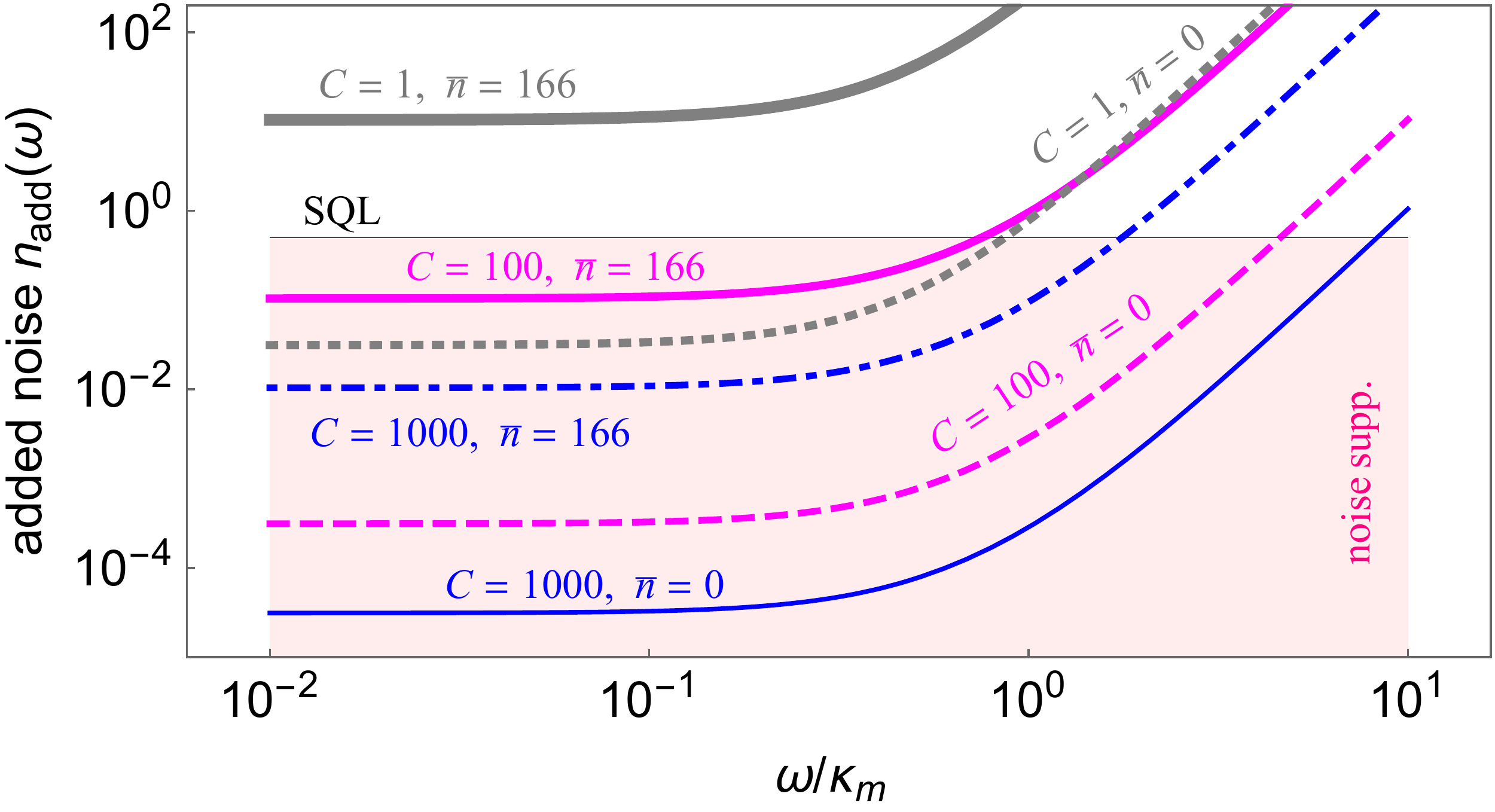}
\caption{(Color online) The effect of thermal noises, temperature, on the added noise of magnetic field measurement. All curves have been plotted beyond the RWA  and for \textit{optimized} zero detuning. 
The blue solid, magenta dashed, and gray dotted curves correspond to zero temperature, $\bar{n}=0$, respectively for $C=1000,100,1$. Moreover, the blue dot-dashed, magenta solid and gray solid curves are referred to room temperature corresponding to $\bar{n}=166$ respectively for $C=1000,100,1$. For obtaining thermal noises, we have chosen $\omega_{a(m)}/2\pi=37.5~\rm {GHz}$ as an achievable bare frequency in the coupled system beyond the RWA \cite{17}.}
\label{Fig4}
\end{figure}

Finally, to understand the roles of temperature or thermal noise on the added noise of magnetic field measurement, in Fig. \ref{Fig4} we have plotted the added noise of magnetic field measurement $n_{\mathrm{add}}(\omega)$ versus the normalized frequency $\omega/ \kappa_m$ at zero detuning, $\Delta=0$, and for two different temperatures, nearly zero and room temperature. 
As is clearly seen and expected from Eq. (\ref{34}), lower temperature corresponding to the smaller thermal noises enables us to more noise suppression.
Interestingly, our sensor has capability to work sensitively even at room temperature in a wide range of frequency (in on  and off resonance) by controlling the cooperativity to be large enough,  like $C=1000$. As is evident, at room temperature in off-resonance frequencies, the noise suppression is still possible by controlling the cooperativity.

It worth remembering  again that beyond the RWA, our proposed magnetic sensor can work sensitively even at room temperature in a wide range of frequency $ \omega \lesssim \kappa_m $, so that simultaneously the noise and signal can be suppressed and amplified, respectively, to obtain a desired precise measurement.

\subsection{\label{secB} Under the RWA}

Considering the system with coupling strength, $g_0$, much smaller than the bare frequencies of the uncoupled system $(g_0\ll \omega_a\simeq \omega_m)$ , one can use the RWA to investigate the system dynamics. Following the same method as Sec. \ref{secA}, the magnonic response and the added noise of measurement under the RWA  and in on-resonance frequency can be obtained as follows (to derive the Hamiltonian and the system dynamics, see appendix)
\begin{eqnarray}
R^{\prime}_m(\omega=0)\vert_{\Delta=0}&=&\frac{4C^{\prime}}{(1+C^{\prime})^2}, \label{37}\\
n^{\prime}_{\mathrm{add}}(\omega=0)\vert _{\Delta=0}&=&(\bar{n}_a+\frac{1}{2})\frac{(1-C^{\prime})^2}{4C^{\prime}}, \label{38}
\end{eqnarray}
where $R^{\prime}_m$, $n^{\prime}_{\mathrm{add}}$, and $C^{\prime}\equiv \frac{4g_0^2}{\kappa_a\kappa_m}\big \vert ^{\mathrm{RWA}}$, respectively, indicate the magnonic response, the added noise, and the electromagnonic cooperativity under the RWA. Note that the above equations are written in the \textit{optimized} zero detuning. Furthermore, it should be noted that under the RWA, in which we use no time modulation, controllability on the coupling strength $ g_0 $ decreases and can be tuned only by controlling the magnetic field inside the cavity. It means that the controllability on the system via the cooperativity under the RWA decreases compared to the system beyond the RWA where the cooperativity can be manipulated easily through $ g=g_0 \mathcal{E} $ via the modulation amplitude $  \mathcal{E} $.

As is evidenced from Eqs. (\ref{37}) and (\ref{38}), unlike beyond the RWA, here, under the RWA the signal response in on-resonance frequency never can be amplified such that at the best condition it can be reached to the unity for the \textit{optimized} cooperativity $ \mathcal{C}'_{\rm opt}=1 $. The added noise on-resonance can be still suppressed perfectly for this \textit{optimized} cooperativity $ \mathcal{C}'_{\rm opt}=1 $. In other words, the proposed magnetic sensor under the RWA never can amplify the signal response but still suppresses the noise below the SQL.

\begin{figure}
\includegraphics[width=8cm]{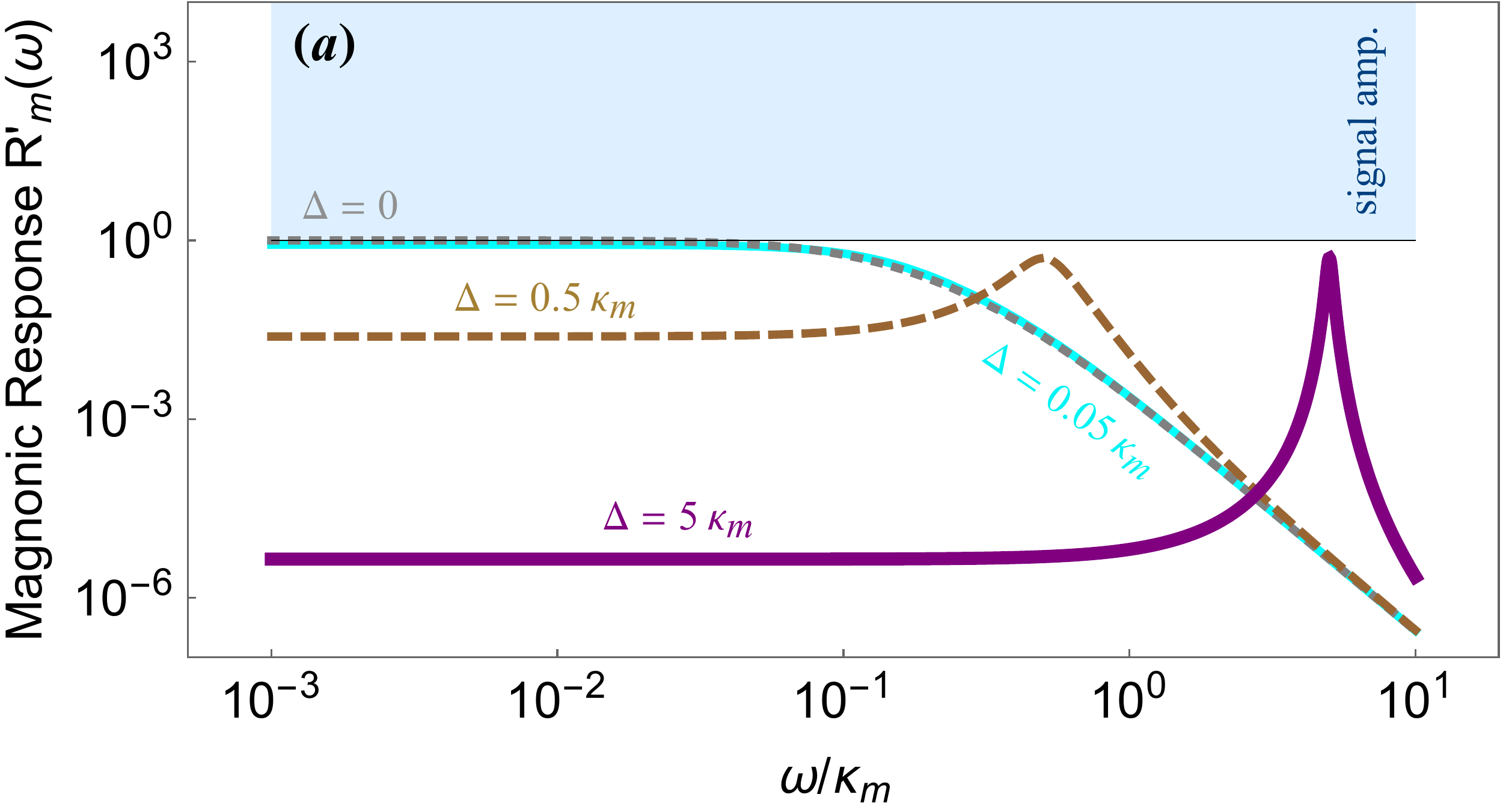}
\hspace{10mm}
\includegraphics[width=8cm]{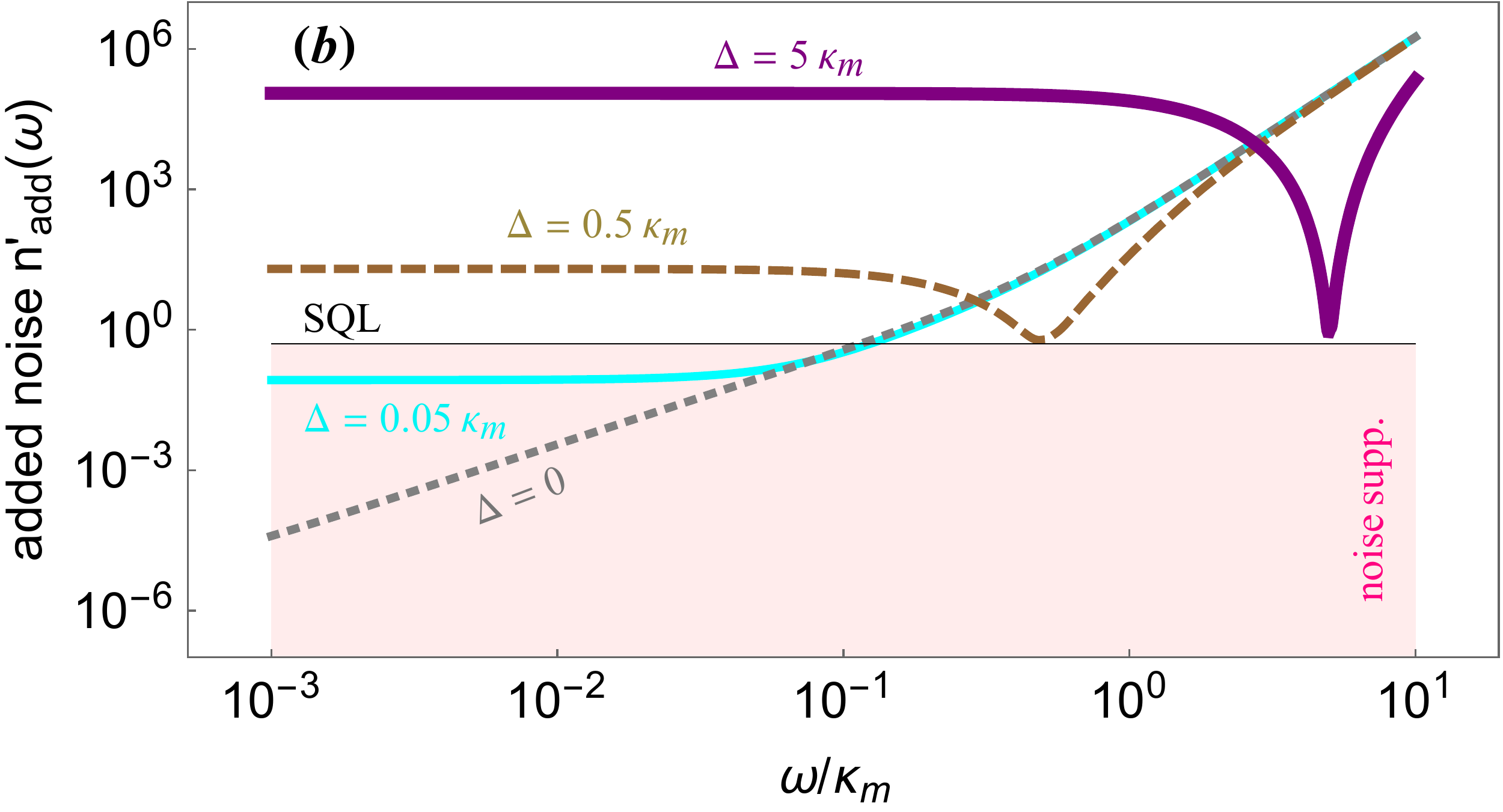}
\caption{(Color online) (a) The magnonic response to the input signal, $R'_m(\omega)$, and (b) the added noise of magnetic field measurement, $n'_{\mathrm{add}}(\omega)$ under the RWA, vs the normalized frequency $\omega/\kappa_m$   for the different detunings in \textit{optimized} cooperativity $C'=1$. The gray dotted, thin cyan solid, brown dashed, and thick purple solid curves, respectively, referred to different detunings $\Delta=0,0.05\kappa_m, 0.5\kappa_m, 5\kappa_m$. Here, we have assumed $\bar{n}\simeq 0$, corresponding to the zero temperature.}
\label{Fig5}
\end{figure}

Similar to Sec. \ref{secA}, in order to see how different values of detuning affect the signal amplification and noise suppression beyond the RWA, in Fig. \ref{Fig5} we have plotted the magnonic response of the cavity output phase spectrum $R'_m(\omega)$ [Fig.  \ref{Fig5}(a)] and the added noise of magnetic field measurement $n'_{add}(\omega)$ [Fig. \ref{Fig5}(b)] with respect to the normalized frequency $\omega / \kappa_m$ for different values of detuning. 
The obtained results for the zero temperature and \textit{optimized} cooperativity $C^{\prime}=1$ show that increasing the detuning, nonzero detuning, leads to signal response decrement. Although for small detunings, like $\Delta=0.05\kappa_m$, the magnonic response is nearly the same as for zero detuning,   Fig. \ref{Fig5}(b) shows the strongest noise suppression occurs for the \textit{optimized} detuning $\Delta=0$. 
Also, to have a precise measurement one should control the detuning and cooperativity to be their optimized values to achieve the maximum response, $ {R'}_m^{\rm max}=1 $, and lower noise in a pretty wide range of frequency $(\omega \le\ 0.1\kappa_m)$.


\begin{figure}
\includegraphics[width=8cm]{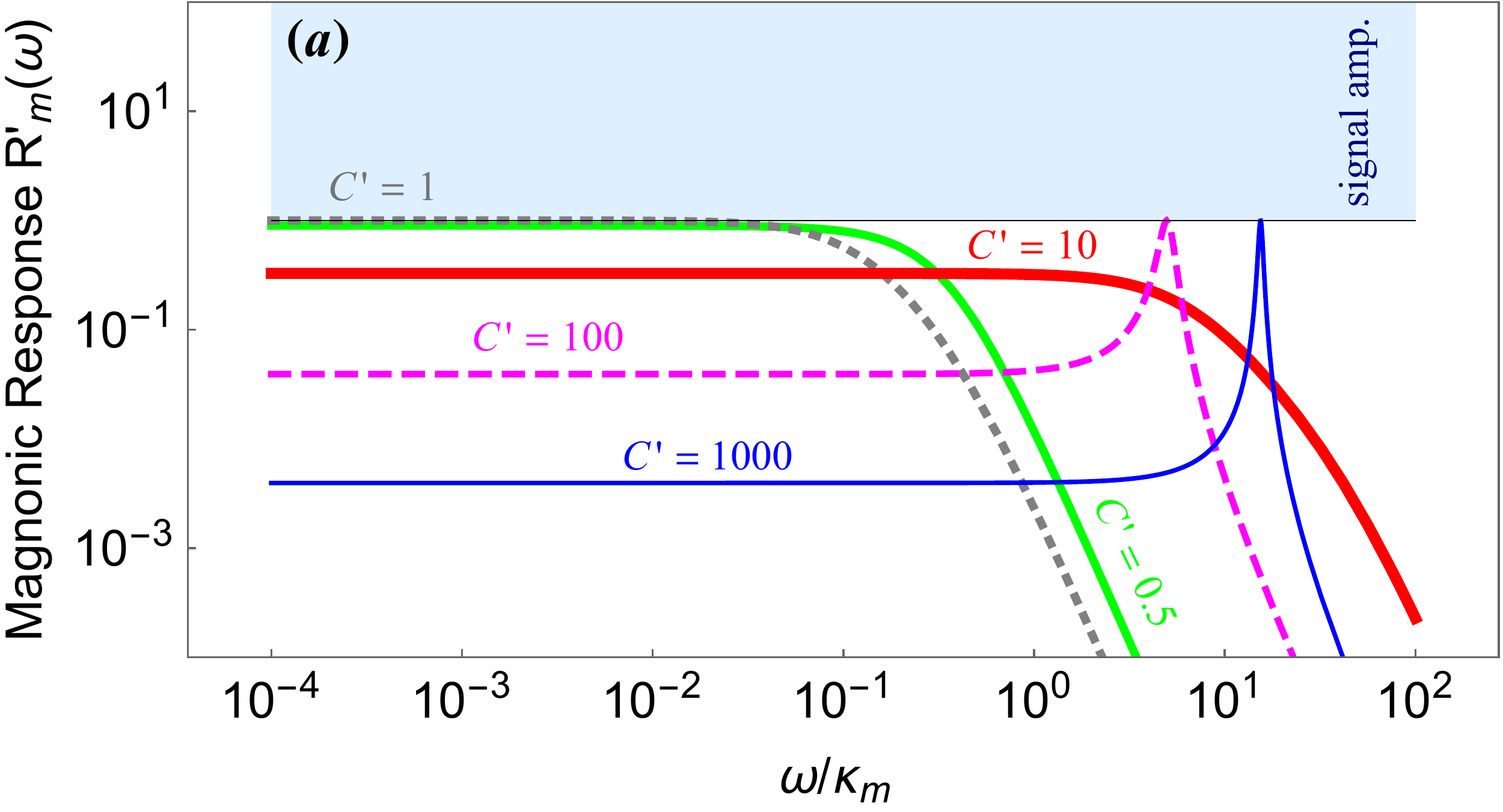}
\hspace{10mm}
\includegraphics[width=8cm]{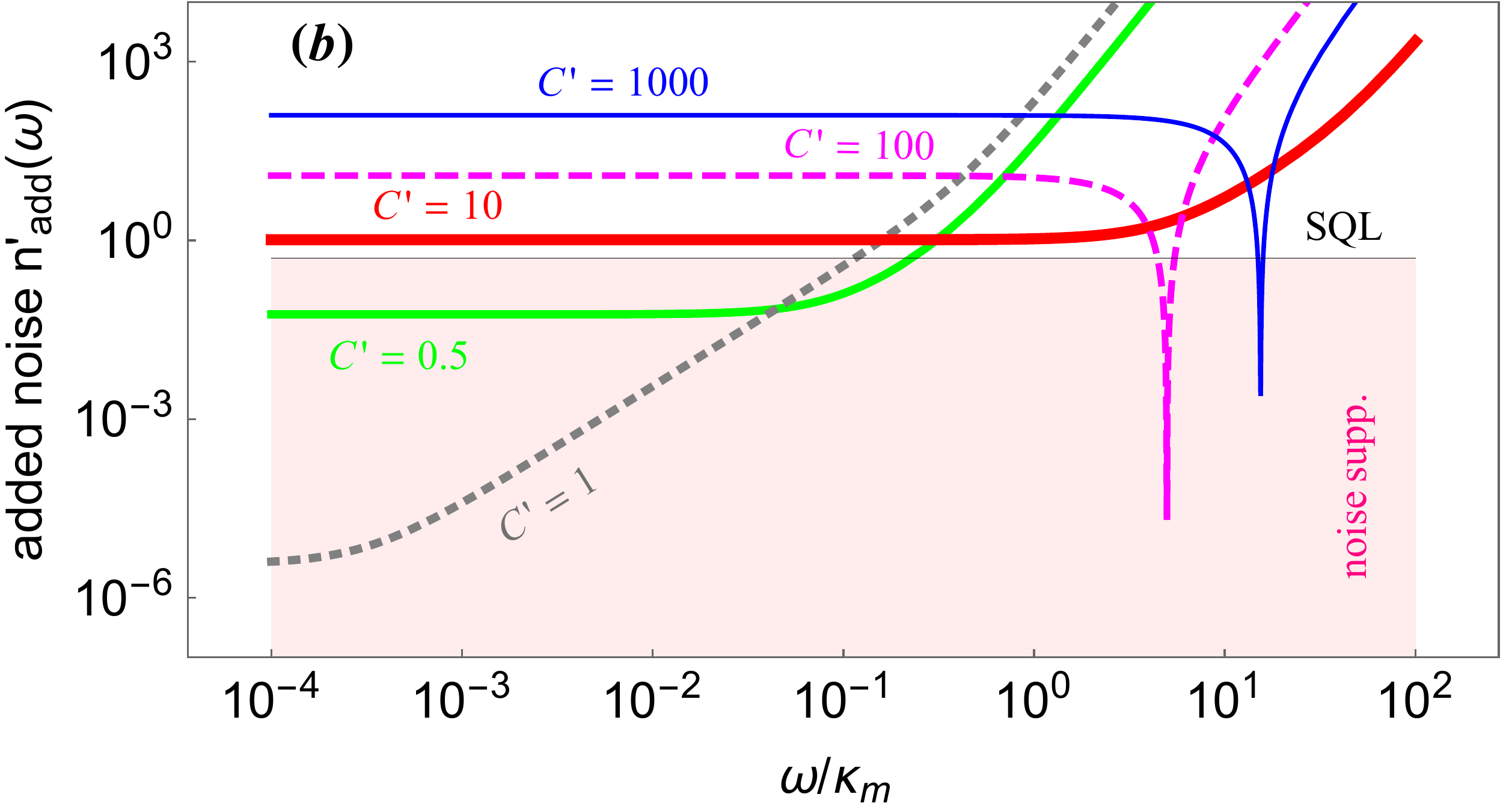}
\caption{(Color online)(a) The magnonic response to the input signal, $R'_m(\omega)$, and (b) the added noise of magnetic field measurement , $n'_{\mathcal{add}}(\omega)$  under the RWA, vs the normalized frequency $\omega/\kappa_m$ for the different cooperativities. All curves have been plotted for \textit{optimized} zero detuning, $\Delta=0$. The thin blue solid, magenta dashed, thick red solid, gray dotted, and green solid curves are respectively referred to  $C'=1000, 100, 10, 1, 0.5$. Here, we have assumed $\bar{n}\simeq0$, corresponding to the zero temperature.}
\label{Fig6}
\end{figure}

To follow the influences of cooperativity on signal amplification and noise reduction, by selecting the \textit{optimized} detuning, $\Delta=0$, and considering zero temperature, $\bar{n}=0$,  in Fig.  \ref{Fig6}  we have plotted the magnonic response of the cavity output phase spectrum $R'_m(\omega)$ [Fig.\ref{Fig6}(a)] and the added noise $n'_{add}(\omega)$ [Fig. \ref{Fig6}(b)] versus the normalized frequency $\omega/\kappa_m$ for different values of cooperativity.
As is clearly evident and expected from Eqs.~(\ref{37}) and (\ref{38}), the optimized value of cooperativity is $C'=1$,  which simultaneously leads to the maximum value of magnonic response, $R'_m(\omega)=1$, and strong noise suppression over a wide range of frequency $(\omega \le\ 0.1\kappa_m)$. 
For smaller cooperativities like $C^{\prime}=0.5$, we can achieve the signal response near to its maximum, but the noise suppression decreases with respect to the \textit{optimized} cooperativity. 
Moreover, for higher cooperativity than its optimized value, for example, $C^{\prime}=100$, simultaneous signal amplification and noise suppression is impossible in a moderate range of frequencies except for a very short range. The situation is worse for very high cooperativities, for example, $C^{\prime}=1000$.

\begin{figure}
\includegraphics[width=8cm]{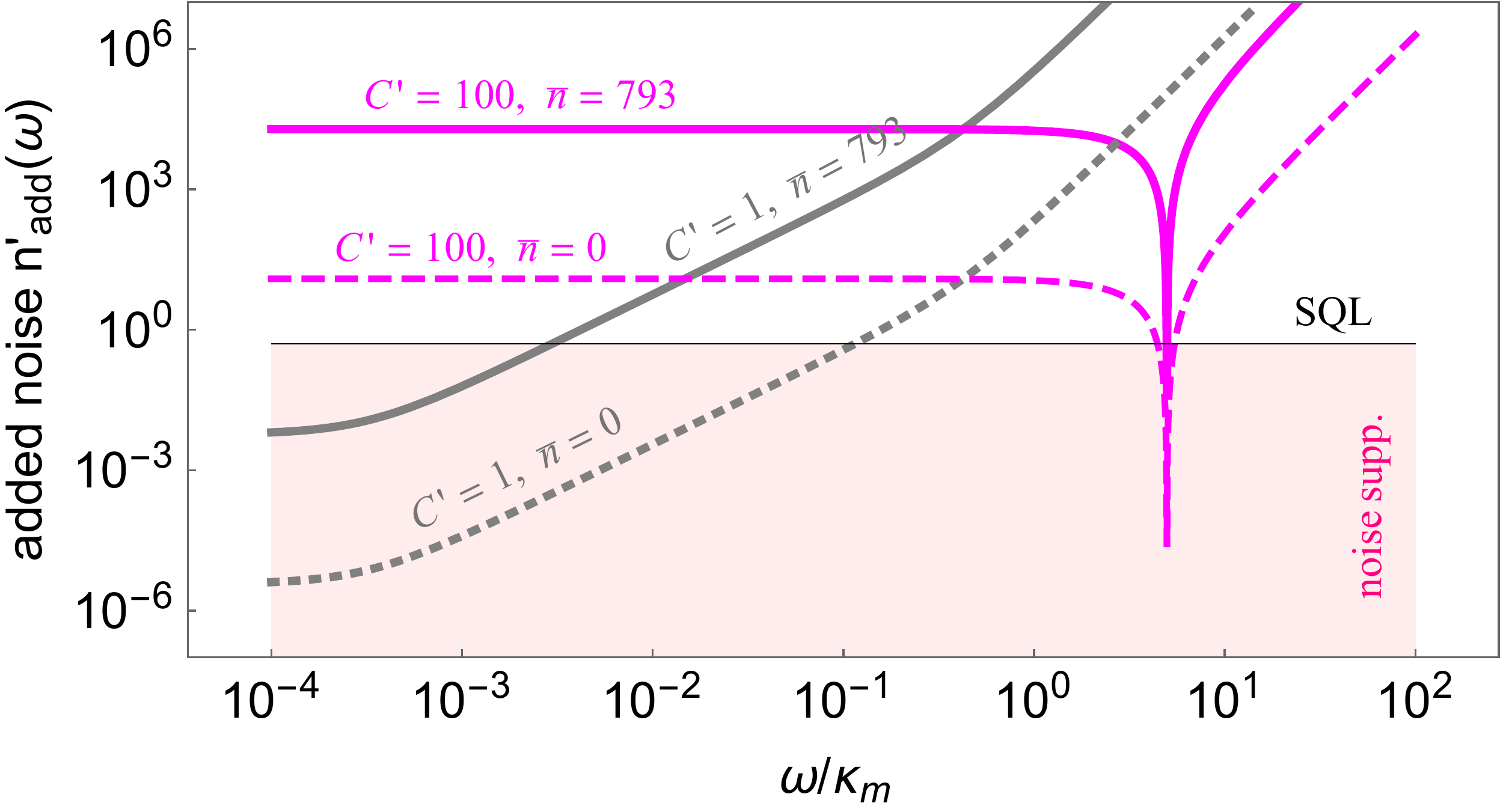}
\caption{(Color online)The effect of thermal noise, temperature, on the added noise of magnetic field measurement under the RWA. All curves have been plotted for the zero detuning, $\Delta=0$. 
The magenta dashed and gray dotted curves are referred to zero temperature, $ \bar n=0 $, respectively for $ C'=100 $ and $ C'=1 $, while the magenta and gray solid curves are referred to room temperature corresponding to $ \bar n=793 $ respectively for $ C'=100 $ and $ C'=1 $. For obtaining thermal noise, we have considered $\omega_{a(m)}/2\pi=7.875~\rm{GHz}$ as an experimental reported bare frequencies in the coupled system  under the RWA \cite{17}.}
\label{Fig7}
\end{figure}

To see the effect of temperature on added noise suppression under the RWA, in Fig. \ref{Fig7}  we have plotted $n'_{\mathrm{add}}(\omega)$ versus the normalized frequency $\omega/\kappa_m$ for zero and room temperature. Similar to the results obtained beyond the RWA, by choosing the \textit{optimized} parameters ($\Delta=0$, and $C^{\prime}=1$) the noise suppression is still possible even at room temperature and over a pretty wide range of frequency ($ \omega \sim 0.1 \kappa_m $) in on- and off-resonance frequencies.

To conclude, under the RWA, no signal amplification occurs, i.e., $ {R'}_m^{\rm max}=1 $. However, in the \textit{optimized} system parameters ($\Delta=0$ and $C^{\prime}=1$), the noise suppression below the SQL is still possible in a pretty wide range $ \omega \sim 0.1 \kappa_m $ at room temperature. Therefore, one can infer that the system beyond the RWA is the optimized case for operational the proposed magnetic sensor. To illustrate this, in the next subsection we compare the signal response and noise suppression of the both cases (beyond and under the RWA) for the fixed \textit{optimized} achievable parameters.


\subsection{\label{secC} Optimized operational regime for the proposed sensor:  Comparison between system beyond the RWA and under the RWA  }

First, compare Figs.~\ref{Fig3} and \ref{Fig6}, which show the results respectively beyond the RWA and under the RWA at zero temperature and \textit{optimized} zero detuning. 
Beyond the RWA, all values of cooperativity can lead to simultaneous signal amplification and noise suppression in a wide frequency range ($ \omega < \kappa_m$), while under the RWA, only by selecting $C'=1$  in a short frequency range ($\omega< 0.1\kappa_m$) can one  achieve the strong noise suppression and the maximum value of magnonic response, ${R'}_m^{\rm max}=1$, which means no signal amplification occurs, in contrast to the obtained results beyond the RWA.
This can be interpreted by the role of counter-rotating terms (CRTs) of the microwave photon--magnon interaction Hamiltonian, which are omitted under the RWA. This can be seen as symmetry breaking in the interaction term. Thus, CRTs are responsible for high signal amplification of the system beyond the RWA, which yields to precise measurement.
In fact, under the RWA, one magnon's worth of input external magnetic field produces at most one photon's worth of output light. Therefore, in this case  although the quantum noise can be suppressed below the SQL, the response cannot be amplified and the system provides only transduction of the external magnetic field \cite{3,5,Review of force sensing}. In contrast, beyond the RWA, i.e., in the presence of the nonlinear terms, one magnon's worth of input external magnetic field can produce more than one photon's worth of output light. Consequently, not only can the system suppress the quantum noise below the SQL, but it can behave as a quantum linear \textit{amplifier} with controllable gain \cite{3,5,Review of force sensing}.



\begin{figure}
\includegraphics[width=8cm]{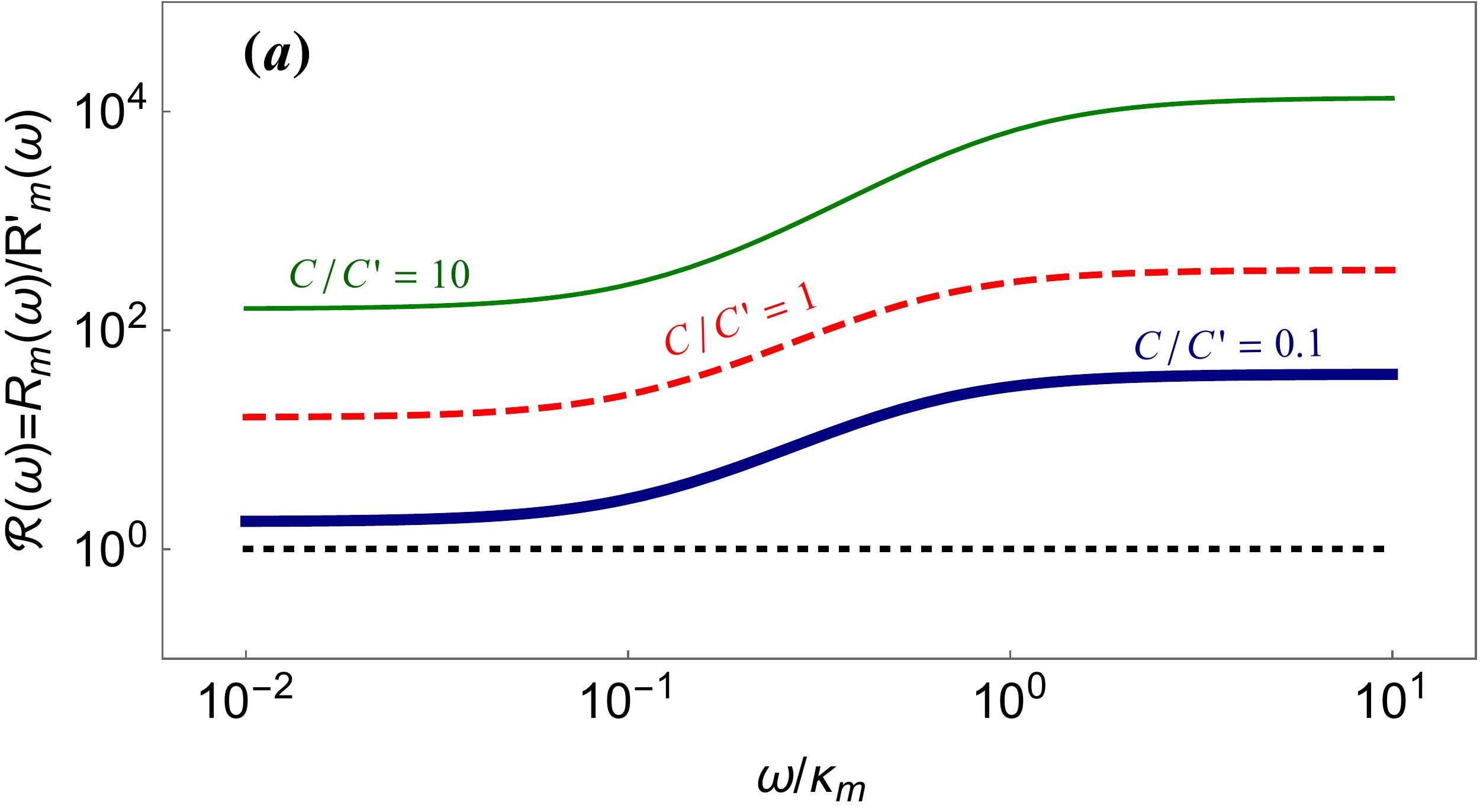}
\hspace{10mm}
\includegraphics[width=8cm]{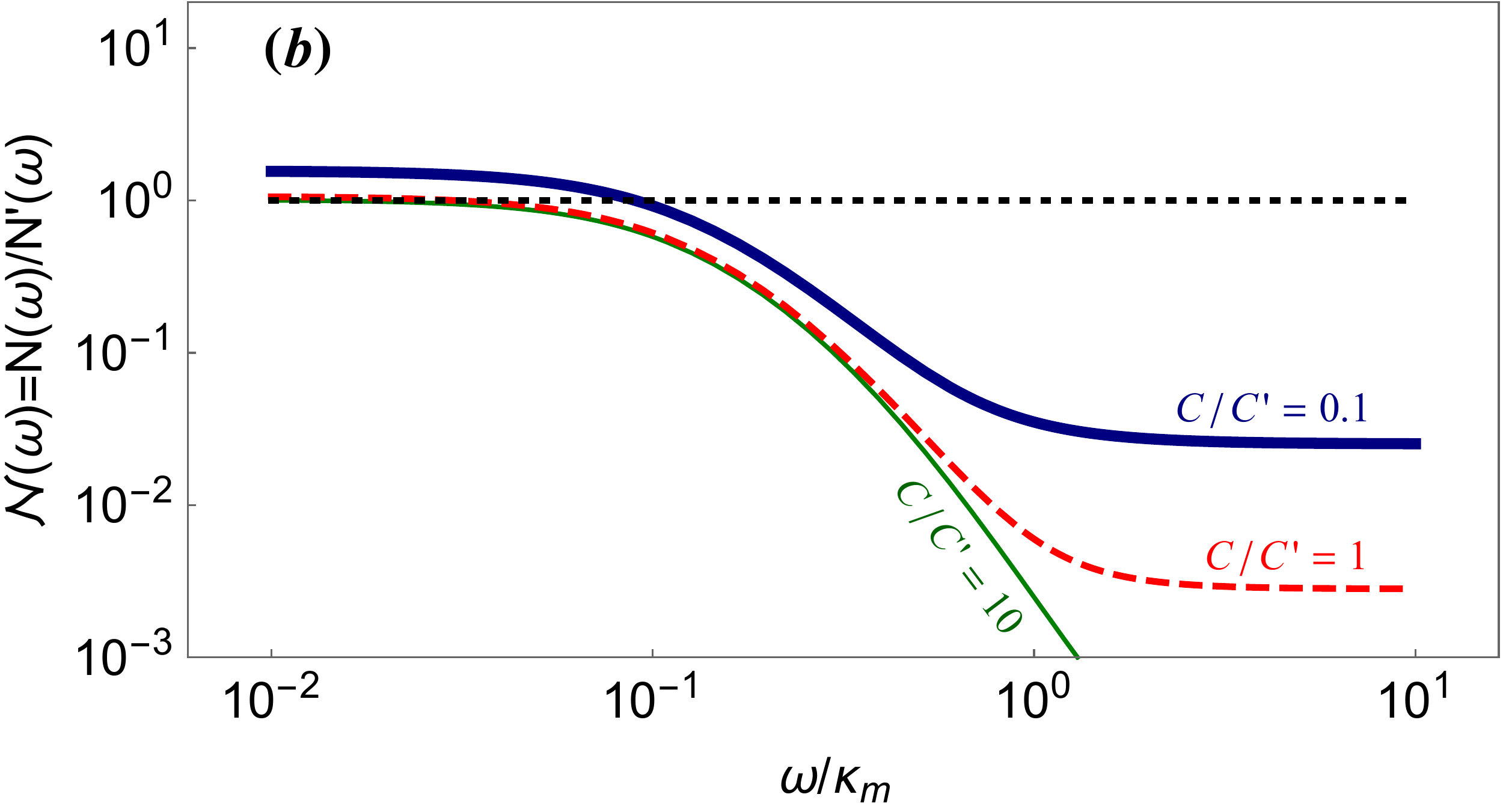}
\caption{(Color online)(a) The ratio of the magnonic response, $\mathcal{R}(\omega)=R_m(\omega)/R'_m(\omega)$, and (b) the  noise of magnetic measurement ratio, $\mathcal{N}(\omega)=N(\omega)/N'(\omega)$, versus the normalized frequency, $\omega/\kappa_m$. The dark blue-solid-thick, red-dashed, and dark green-solid-thin curves respectively correspond to cooperativities ratios $C/C'=0.1,1.10$. Here, we have considered the \textit{optimized} zero detuning and also zero-temperature for both cases. All parameters have been chosen so the system to be in the optimized sensing regime where the added nosies are suppressed and signal responses are amplified.}
\label{Fig8}
\end{figure}

To better compare, in Fig.~\ref{Fig8} we have plotted the ratio of the magnonic response beyond the RWA to the corresponding one under the RWA, $\mathcal{R}(\omega)=\frac{R_m(\omega)}{R'_m(\omega)}$ [Fig. \ref{Fig8}(a)] as well as the ratio of their added noise, $\mathcal{N}(\omega)=\frac{N(\omega)}{N'(\omega)}$  [Fig. \ref{Fig8}(b)], versus the normalized frequency, $\omega/\kappa_m$, for different ratios of cooperativities, $C/C'=10,1,0.1$. 
Here, we have considered $N(\omega)=[(\bar{n}+\frac{1}{2})+n_{\mathrm{add}}(\omega)]$ and $N'(\omega)=[(\bar{n}+\frac{1}{2})+n'_{\mathcal{add}}(\omega)]$ as the induced total noise  to the magnetic field measurement beyond the RWA and under the RWA, respectively. 
For true comparison, we have set parameters for zero temperature and \textit{optimized} zero detuning in both cases while we have changed cooperativities 
in their optimized values range such that the cooperativities ratio can be tuned from small to large values ($0.1$ to $10$). We note that the optimum cooperativity under the RWA is $C'=1$, and then is always fixed and cannot be changed for plotted curves in Fig.  \ref{Fig8}. However, as has been shown, beyond the RWA one can set all values for the cooperativity since for all values simultaneous signal amplification and noise suppression are possible. Thus, we have considered three different cooperativities set as ($C'=1, C=10$), ($C'=1, C=1$), and ($C'=1, C=0.1$).

As  illustrated in Fig. \ref{Fig8}(a), the magnonic response to the external input in the \textit{optimized} parameters set for the system beyond the RWA is always larger than the system under the RWA, notably for higher ratio of cooperativities, $C/C'$. Also, higher cooperativity ratios correspond to the better operation of the sensor, i.e., more signal amplification for the system beyond the RWA compared to that under the RWA. Another advantage of the system beyond the RWA is its higher response in off resonance frequencies. It means that in far off-resonance, for example, $\omega > 0.1\kappa_m  $ for considered parameters in Fig.~(\ref{Fig8}), the proposed sensor beyond the RWA can work better than the sensor under the RWA. On the other hand, one can conclude that the room temperature operation of the proposed sensor beyond the RWA is much better than the RWA especially for far off-resonance frequencies. (Although it is not shown here,  the results of comparison at room temperature are the same as the obtained results at zero temperature.)

Figure \ref{Fig8}(b) shows the ratio of added noises for different values of cooperativity ratio. As is evident, $\mathcal{N}(\omega)=1$ for frequency range $\omega \le \ 0.1\kappa_m$ which can be considered as near on-resonance range. It means that for near on-resonance frequencies the noise suppression in both cases is approximately the same. [Note that near the resonance the added noise of the sensor under the RWA in the \textit{optimized} parameters zero detuning and $ C'=1 $ is exactly zero; see Eq.  \ref{38}.] But, in off-resonance range $\omega > 0.1\kappa_m$, the added noise ratio $\mathcal{N}(\omega)$ decreases which means that for off-resonance range noise suppression of the sensor beyond the RWA is stronger than the sensor under the RWA.

Finally, we conclude that the operation of our proposed sensor beyond the RWA is more appropriate for on-resonance and especially off-resonance quantum magnetometry in a wide frequency range. Furthermore, it should be mentioned that the results for negative detuning are exactly the same as presented results for positive detuning.

\section{\label{sec5} SENSITIVITY, SIGNAL-TO-NOISE RATIO, and experimental discussion}

In this section, we calculate the SNR and also the sensitivity of the system. Beyond the RWA, we can write the output phase quadrature of the cavity field  as 
\begin{eqnarray}\label{39}
P^{\mathrm{out}}_a(\omega)&=&P^{\mathrm{out}}_a(\omega)\vert_{B=0} +i\mathcal{A}(\omega)\tilde{B}_1(\omega)\eta\sqrt{\frac{2}{\kappa_m}}\nonumber\\
&&+\mathcal{B}(\omega)\tilde{B}_2(\omega)\eta \sqrt{\frac{2}{\kappa_m}},\
\end{eqnarray}
where 
\begin{eqnarray}\label{40}
P^{\mathrm{out}}_a(\omega)\vert_{B=0}&=&\mathcal{A}(\omega)X^{\mathrm{in}}_m(\omega)+\mathcal{B}(\omega)P^{\mathrm{in}}_m(\omega)+\mathcal{C}(\omega)X^{\mathrm{in}}_a(\omega)\nonumber\\
&&+\mathcal{D}(\omega)P^{\mathrm{in}}_a(\omega)
\end{eqnarray}
is the contribution of the output phase quadrature in the absence of the external magnetic field (that we are going to sense). $\tilde{B}_1(\omega)$ and $\tilde{B}_2(\omega)$ are the contribution of the external magnetic field and are defined as follows
\begin{eqnarray}
\tilde{B}_1(\omega)&=&\frac{1}{2}[B(\omega+\omega_d)-B(\omega-\omega_d)], \label{41}\\
\tilde{B}_2(\omega)&=&\frac{1}{2}[B(\omega+\omega_d)+B(\omega-\omega_d)].\label{42}
\end{eqnarray}

In order to calculate the sensitivity of the device to the external magnetic field and the SNR, we follow the method used in Ref.~\cite{5}. To this end, we define the magnetic field operators as 
\begin{eqnarray}
\delta \tilde{B}_1(\omega)&=&\frac{P^{\mathrm{out}}_a(\omega)}{\partial P^{\mathrm{out}}_a(\omega)/\partial \tilde{B}_1(\omega)}=\delta \hat{N}_1+\tilde{B}_1(\omega), \label{43}\\
\delta \tilde{B}_2(\omega)&=&\frac{P^{\mathrm{out}}_a(\omega)}{\partial P^{\mathrm{out}}_a(\omega)/\partial \tilde{B}_2(\omega)}=\delta \hat{N}_2+\tilde{B}_2(\omega), \label{44}
\end{eqnarray}
in which $\delta\hat{N}_1(\omega)$ and $\delta\hat{N}_2(\omega)$ are the noise operators and affect the system through two different channels $\hat{X}^{\mathrm{in}}_m(\omega)$ and $\hat{P}^{\mathrm{in}}_m(\omega)$ independently and are defined as follows:
\begin{eqnarray}
\delta \hat{N}_1(\omega)&=&\frac{\mathcal{A}(\omega) \hat{X}^{\mathrm{in}}_m+\mathcal{B}(\omega) \hat{P}^{\mathrm{in}}_m+\mathcal{C}(\omega) \hat{X}^{\mathrm{in}}_a+\mathcal{D}(\omega) \hat{P}^{\mathrm{in}}_a}{i\mathcal{A}(\omega)\eta \sqrt{\frac{2}{\kappa_m}}}, \label{45}\\
\delta \hat{N}_2(\omega)&=&\frac{\mathcal{A}(\omega) \hat{X}^{\mathrm{in}}_m+\mathcal{B}(\omega) \hat{P}^{\mathrm{in}}_m+\mathcal{C}(\omega) \hat{X}^{\mathrm{in}}_a+\mathcal{D}(\omega) \hat{P}^{\mathrm{in}}_a}{\mathcal{B}(\omega)\eta \sqrt{\frac{2}{\kappa_m}}}. \label{46}
\end{eqnarray}

Considering the probability amplitudes of the noise injection through different channels, the total noise operator can be written as follows
\begin{equation}\label{47}
\delta \hat{N}(\omega)=P_1\delta\hat{N}_1(\omega)+P_2\delta\hat{N}_2(\omega),
\end{equation}
where the probability amplitudes of noise injection ($P_1$ and $P_2$ ) through $\hat{X}^{\mathrm{in}}_m$ and $\hat{P}^{\mathrm{in}}_m$ are given by
\begin{eqnarray}
P_1&=&\frac{\mathcal{A}(\omega)}{\sqrt{\vert \mathcal{A}(\omega)\vert ^2+\vert \mathcal{B}(\omega)\vert ^2}}, \label{48}\\
P_2&=&\frac{\mathcal{B}(\omega)}{\sqrt{\vert \mathcal{A}(\omega)\vert ^2+\vert \mathcal{B}(\omega)\vert ^2}}. \label{49}
\end{eqnarray}
Therefore, the total noise operator is obtained as follows:
\begin{eqnarray}\label{50}
\delta  \hat{N}(\omega)&=&\frac{(1-i)}{\eta}\sqrt{\frac{\kappa_m}{2}}\nonumber\\
&&(\frac{\mathcal{A}(\omega) \hat{X}^{\mathrm{in}}_m+\mathcal{B}(\omega) \hat{P}^{\mathrm{in}}_m+\mathcal{C}(\omega) \hat{X}^{\mathrm{in}}_a+\mathcal{D}(\omega) \hat{P}^{\mathrm{in}}_a}{\sqrt{\vert \mathcal{A}(\omega)\vert ^2+\vert \mathcal{B}(\omega)\vert ^2}}), \nonumber\\
\end{eqnarray}
which yields to noise power spectrum as
\begin{eqnarray}\label{51}
S_N(\omega)=\frac{1}{2}\langle\delta \hat{N}(\omega)\delta \hat{N}(\omega)^{\dagger}\rangle, 
\end{eqnarray}
which can be simplified as
\begin{equation}\label{52}
S_N(\omega)=\frac{\kappa_m}{\eta^2}[(\bar{n}_m+\frac{1}{2})+n_{\mathrm{add}}(\omega)].
\end{equation}
SNR, the ratio of the signal to the variance of the noise, can be written as
\begin{equation}\label{53}
r(\omega)=\frac{\vert \tilde{B}(\omega) \vert}{\sqrt{S_N(\omega)}}=\frac{\vert \tilde{B}(\omega) \vert}{\frac{\sqrt{\kappa_m}}{\eta}[(\bar{n}_m+\frac{1}{2})+n_{\mathrm{add}}(\omega)]^{\frac{1}{2}}}.
\end{equation}
The sensitivity or the minimum detectable input of the device is the minimum magnitude of the input signal required to produce an output with $r(\omega)=1$. Therefore, sensitivity of the system beyond the RWA, $\mathcal{S}(\omega)=\sqrt{S_N(\omega)}$, is simplified as 
\begin{equation}\label{54}
\mathcal{S}(\omega)^{\rm bRWA}=\frac{\sqrt{\kappa_m}}{\eta}[(\bar{n}_m+\frac{1}{2})+n_{\mathrm{add}}(\omega)]^{\frac{1}{2}}.
\end{equation}

The feasible experimental parameters beyond the RWA \cite{17} are $\omega_m/2\pi \simeq \omega_a/2\pi=37.5 ~\rm GHz$,  $g_0/2\pi= 2.5 ~\rm GHz$, $\kappa_m/2\pi=15~ \rm MHz$, and $\kappa_a/2\pi=33 ~\rm MHz$. This coupled microwave photon--magnon system that yields to a ratio of $g_0/\omega_{a(m)}=0.067 $ has  been realized by placing a YIG sphere (which spin and orbital angular momentum of its unit cell are respectively $s=\frac{5}{2}$ and $l=0$ \cite{26}) with diameter of  $2.5 ~\rm mm$ inside a microwave cavity with dimension   $(7.0\times5.0\times3.2)~ \rm mm^3$ and with a bias magnetic field as $B_b=1.34~ \rm{T}$, which is exerted to the YIG sphere.
 Therefore, the obtained electromagnonic cooperativity  is $C= \frac{4g^2}{\kappa_a\kappa_m} \simeq 5 \mathcal{E}^2\times 10^4$ 
 in which $\mathcal{E}$ can be controlled so that the condition $g=g_0\mathcal{E} \ll \ \Omega  (\Omega=2\omega_L)$ to be satisfied. 
It is worth  pointing out that in practice, for controlling the $\mathcal{E}$ and therefore the magnon--photon coupling strength one can use a Josephson parametric amplifier (JPA).  By changing the flux bias through the SQUID loop in the JPA, the inductance of the SQUID is modulated. Accordingly, the modes in the resonator can be deformed. By placing a YIG sphere where the magnetic field is maximum and controlling the modulation amplitude of magnetic field, one can change $\mathcal{E}$ and hence $g$ so that $g \ll \ \Omega$ being valid.
 In this manner, control of the cooperativity beyond the RWA is possible via controlling the time modulation amplitude $\mathcal{E}$ which leads to the different cooperativities considered in the figures in Sec. \ref{secA}.
The gyromagnetic ratio is $\gamma/2\pi= 28 ~\frac{ \rm GHz}{\rm{T}}$ \cite{17}, and the spin density of YIG is $\rho_s=4.22\times 10^{27}~ \rm m^{-3}$ which yield  the total number of spins $N=3.5\times 10^{19}$ \cite{17}. 
Now, by considering these feasible experimental parameters in the \textit{optimized} detuning, $ \Delta=0 $, the sensitivity of the theoretical proposed sensor near to on-resonance frequency beyond the RWA corresponding to USC or DSC regime is obtained as $\mathcal{S}_0^{\rm bRWA}=5.90\times 10^{-18} \frac{\rm{T}}{\sqrt{\rm Hz}}$ and $\mathcal{S}_{\rm room}^{\rm bRWA}=1.08\times 10^{-16} \frac{\rm{T}}{\sqrt{\rm Hz}}$, respectively, at zero and room temperature. 
Moreover, by considering the feasible experimental frequency, $\omega_{a(m)}$, the thermal noises at zero and room temperature are respectively  $\bar{n}_{0}^{\rm bRWA}=0$ and $\bar{n}_{\rm room}^{\rm bRWA}=166$. 
As is evident from Eq. (\ref{54}), $ \sqrt{\kappa_m}/\eta $ (with $ \eta:= \frac{\gamma}{2}\sqrt{5N} $) plays the key role for determination of the sensitivity in the optimized condition. Thus, by increasing the number of spins in the YIG sphere through increasing its diameter during the fabrication process, and decreasing the magnon decay rate, $\kappa_m$,  one may achieve to a more sensitive magnetometer beyond the RWA.



Following the same  procedure, one can obtain the sensitivity of our sensor under the RWA as follows:
\begin{equation}\label{55} 
\mathcal{S}^{\rm{RWA}}(\omega)=\frac{\sqrt{\kappa_m}}{\eta}[(\bar{n}_m+\frac{1}{2})+n^{\prime}_{\mathrm{add}}(\omega)]^{\frac{1}{2}}.
\end{equation}

The experimental parameters corresponding to the RWA condition ($g_0\ll \ \omega_{a(m)}$) are $\omega_m/2\pi \simeq\omega_a/2\pi=7.875 \rm GHz$, $g_0/2\pi=3.1 \rm MHz$, $\kappa_m/2\pi=19 \rm MHz$ and $\kappa_a/2\pi=2.09 \rm MHz$ which has been realized by placing a YIG sphere with diameter of $0.36 \rm mm$ inside a microwave cavity with an inner dimension $(43.0\times 21.0\times 9.6) \rm {mm}^3$ and also applying a bias magnetic field  $B_b=281 \rm{mT}$ to the YIG sphere \cite{17}. These experimental parameters lead to cooperativity near to the \textit{optimized} value for magnetic sensing under the RWA, $C'=\frac{4g_0^2}{\kappa_m\kappa_a}\big \vert ^{\rm{RWA}}= 0.97 \simeq 1$. 
Here, by a $0.36 \rm mm$ diameter YIG sphere, the total number of spins is $N=1.031 \times 10^{17}$ \cite{17}, which leads to a smaller coupling strenth, $g_0/2\pi=3.1 \rm MHz$, compared to the previous example with a $2.5mm$ diameter YIG sphere leading to the total spin number and coupling strength respectively as    $N=3.5\times 10^{19}$   and   $g_0/2\pi= 2.5 \rm GHz$. Therefore, the sensitivity of our theoretical proposed magnetometer under the RWA near to on-resonance frequency is obtained as $\mathcal{S}_0^{\rm {RWA}}= 1.22\times 10^{-16}\frac{\rm{T}}{\sqrt{ \rm Hz}}$ and $\mathcal{S}_{\rm room}^{\rm {RWA}}=4.87 \times 10^{-15} \frac{\rm{T}}{\sqrt{\rm Hz}}$ at zero and room temperature, respectively. 
Under the RWA, the thermal noises corresponding to the zero  and room temperature are, respectively, $\bar{n}_0^{\rm {RWA}}=0$ and $\bar{n}_{\rm {room}}^{\rm {RWA}}=793$.
Similar to the sensor beyond the RWA,   increasing  the number of spins in the YIG, $N$, by choosing a bigger sphere and decreasing the magnon decay rate, $\kappa_m$,  lead to  more sensitivity.

Note that the bias magnetic field $ B_b $ is usually a static strong background field, for example, $B_b= 1.34 \rm{T}$ and $B_b= 281 \rm {mT}$ respectively for the case of beyond and under the RWA, while the signal magnetic field that we would like to detect is a time-dependent magnetic field. Hence, this strong background can be distinguished and filtered using some experimental filtering technique such as Kalman filtering, which is a well-known technique in engineering and recently has been developed to be used in force sensing in cavity optomechanics \cite{kalman filtering}. For example, similar to our proposed scheme, there is a strong static bias field $B_b= 0.4 ~\rm T$ in an experimental cavity magnon polariton-based magnetometer in Ref.~\cite{phasemodulatedmagnetometry}. Nevertheless, the possibility of weak magnetic field measurement by this magnetometer with a sensitivity of the order of $2.0 \frac{\rm {pT}}{\sqrt{\rm Hz}}$ has been reported \cite{phasemodulatedmagnetometry} by using the  background canceling. Thus, this scheme and the used experimental techniques to distinguish the signal and background can guarantee that our proposed theoretical scheme with a similar strong bias magnetic field $ B_b=1.34 \rm T$ ($ B_b=281 \rm {mT}$) can similarly measure weak magnetic field with sensitivity in the order of $1.08 \times 10^{-16} \frac{\rm T}{\sqrt{\rm Hz}}$   ($4.87 \times 10^{-15} \frac{\rm T}{\sqrt{\rm{Hz}}}$) and $5.90\times 10^{-18} \frac{\rm T}{\sqrt{\rm {Hz}}}$ ($1.22 \times 10^{-16}\frac{\rm T}{\sqrt{\rm {Hz}}}$) beyond (under) the RWA, at room and zero temperature, respectively. 
Note that although the high sensitivities of our theoretical proposed magnetometer are obtained based on the reported experimental parameters \cite{17},  it is still an ideal result which needs to be proved and verified experimentally.


For different set of experimental parameters, the sensitivity of the sensor beyond the RWA is much more than that under the RWA. But, it might be mentioned that by choosing a YIG sphere with the same size considered for the system beyond the RWA or even bigger, one can obtain the sensitivity in the order of the obtained sensitivity beyond the RWA but still a little smaller. Here, the important point is that for the same sensitivity in both cases, the sensor beyond the RWA is still much better for quantum magnetic sensing since it not only suppresses the noise but also \textit{amplifies} the magnetic-signal response as an quantum \textit{amplifier}. 
Therefore, one can conclude that the optimum regime for operation of the proposed sensor is the USC or DSC regime corresponding to system beyond the RWA.

As has been explained in theoretical results of Sec. \ref{sec4}, in this optimized regime quantum noise suppression below the SQL ($n_{\mathrm{add}}^{\mathrm{SQL}}=\frac{1}{2}$), which is required for precise magnetometry, is possible even by weak or intermediate cooperativities. 
But, according to  Eq. (\ref{54}), by suppressing the quantum noise below the SQL, what plays the important role for the sensitivity is the factor $\frac{\sqrt{\kappa_m}}{\eta}$. It implies a low magnon mode dissipation rate and a high total spin number. Since the magnon-photon coupling strength is as $g=\frac{\gamma B_0}{2}\sqrt{2Ns}$, it yields a larger coupling strength $g$, which is necessary for the ultraprecise magnetic field measurement. As is evident, both decrease of the magnon mode decay rate and increase of the total spin number lead to higher cooperativity. To clarify this dependency on the cooperativity, by considering $g=\frac{\gamma B_0}{2}\sqrt{2Ns}$ and also $\eta=\frac{\gamma}{2}\sqrt{2Ns}$, one can easily rewrite the factor $\frac{\sqrt{\kappa_m}}{\eta}$ as 
\begin{eqnarray}\label{56}
\frac{\sqrt{\kappa_m}}{\eta}=\frac{2B_0}{\sqrt{C\kappa_a}},
\end{eqnarray}
in which $B_0$ indicates the magnetic field amplitude of the electromagnetic field. On the other hand, cavity enhancement, which is proportional to the increase of the coupling strength $g$ and decrease of the cavity and magnon mode dissipation rates $\kappa_a$, and $\kappa_m$, is analogous to the increase of cooperativity $C=\frac{4g^2}{\kappa_a \kappa_m}$. Thereby, although the quantum noise suppression is possible even in the weak or moderate cooperativity regimes,   the strong cooperativity regime or \textit{cavity enhancement} is necessary for ultraprecise magnetometry. Surprisingly, these conditions, i.e., ultradense spin ensemble and cavity enhancement, which yield ultrahigh sensitivity, currently have been experimentally realized in Ref ~ \cite{17}, and thus it guarantees our estimate of ultrahigh sensitivity in our proposed magnetometer since we have used these realized experimental values.

Consequently, the \textit{ultrahigh sensitivity} originates from both \textit{ultradense spin ensemble} and \textit{cavity enhancement}. Let us discuss the feasibility of the proposed method in the integrated systems such as coplanar microwave  systems. As we know, ultrahigh cooperativity of electromagnonic interaction based on coplanar microwave waveguides has not been reported yet, but finding or fabricating a ferromagnetic material with an ultradense spin number and low magnon dissipation rate can lead to a cavity-enhanced and consequently  an on-chip integrated ultraprecise magnetometer. 

Let us explain more. In addition to microwave cavities, which can provide photon-magnon interaction conditions as has been reported in Refs.~ \cite{16, 17}, according to some experimental results reported in Refs.~ \cite{Highcooperativity, CPW 2, CPW 3, CPW 4}, the coplanar microwave waveguides can also provide a platform for magnon-photon interaction. Magnon-photon interaction in these systems  is possible not only in the strong coupling regime \cite{CPW 2, CPW 3, CPW 4}, but also (regarding the experimental results of Ref.~\cite{Highcooperativity}) in the ultrastrong coupling regime, where, here, we have shown is necessary for simultaneous signal amplification and noise reduction in order to achieve a highly sensitive magnetometer. However, there are some limitations and challenges with this kind of resonator. The coplanar waveguides used for magnon-photon interaction are mostly made from superconducting materials that work at low temperatures of the order of $\mathrm{mK}$--$\mathrm{K}$, and so there is need for cryogenic cooling \cite{Highcooperativity, CPW 2, CPW 3, CPW 4}. Therefore, these waveguides cannot be used for precise measurement at room temperature, while our proposed system can work precisely at room temperature. Moreover, the ferromagnetic materials used for magnon-photon interaction in coplanar microwave waveguides have a small size, yielding a smaller total spin number, for example, about three orders smaller \cite{Highcooperativity} than our proposed system, which leads to  less cavity enhancement and limits ultraprecise magnetometry. 

For example, the experimental results of Ref.  \cite{Highcooperativity} show the possibility of magnon-photon interaction using a ferromagnet yttrium iron garnet (YIG) doped with gallium (YIG: Ga) on a superconducting coplanar microwave resonator made from Nb. In this case, the resonator structure is patterned into a $100 ~\mathrm{nm}$-thick Nb film deposited onto an intrinsic silicon substrate using photolithography and reactive ion etching. The volume of the commercial YIG: Ga crystal, which is placed in the center of the resonator with a frequency of  $\omega_a/2\pi=5.90 ~\rm{GHz}$ is around $0.7 ~\mathrm{mm^3}$. The total spin number coupled to the resonator is around  $N\simeq 4.5\times 10^{16}$, which is about three orders of magnitude smaller than our proposed macroscopic system, which has a total spin number of about $N=3.5\times 10^{19}$. By considering the total spin number of $N=4.5\times 10^{16}$  in waveguide systems, the magnon-photon coupling strength in this integrated system is around $g/2\pi=450 ~\rm{MHz}$ and magnon dissipation and resonator decay rates are respectively as $\kappa_m/2\pi=50 ~\rm{MHz}$  and $\kappa_a/2\pi=3 ~\rm {MHz}$, which all together lead to a very large cooperativity as $C=\frac{4g^2}{\kappa_a\kappa_m}=5400$. Although it has not been mentioned in Ref.~\cite{Highcooperativity},  by considering $\frac{g}{\omega_{a(m)}} \simeq 0.08$, one says that these experimental results that have been obtained at millikelvin temperatures show the ultrastrong coupling regime of magnon-photon interaction, which based on our proposal is the necessary regime for ultrasensitive magnetometry. In this regime, which guarantees the simultaneous signal amplification and noise reduction, by considering the optimized detuning, then the sensitivity of the integrated microwave system with a cooperativity of about $C=5400$ and a factor $\frac{\sqrt{\kappa_m}}{\eta} \simeq 4.25\times 10^{-16}\frac{\rm T}{\sqrt{\rm Hz}}$ is of the order of $3.01\times 10^{-16} \frac{\rm T}{\sqrt{\rm Hz}}$ near to that of on-resonance frequency at millikelvin temperature. In comparison, based on the reported experimental results of Ref. \cite{17}, our proposed macroscopic magnetometer including a microwave cavity with a dimension of $(7.0\times 5.0 \times 3.2) ~\mathrm{mm^3}$ and a frequency of $\frac{\omega_a}{2\pi}=37.5 ~\rm{GHz}$ and also containing a $8.2~\mathrm{ mm^3}$ YIG sphere with total spin number of $N=3.5\times 10^{19}$ has  photon-magnon coupling strength and magnon and cavity mode dissipation rates respectively of $g/2\pi= 2.5 \rm {GHz}$, $\kappa_m/2\pi= 15\rm{MHz}$   and $\kappa_a/2\pi= 33\rm{MHz}$ , which all lead to a cooperativity of $C \simeq 50000$, which shows a large cavity enhancement that has been experimentally realized \cite{17}. By considering the ultrastrong coupling regime beyond the RWA, the ratio in our case is  $\frac{\sqrt{\kappa_m}}{\eta}\simeq 8.34 \times 10^{-18} \frac{\rm T}{\sqrt{\rm Hz}}$ and then the sensitivity in our case is obtained of the order of $5.90 \times 10^{-18}  \frac{\rm T}{\sqrt{\rm{ Hz}}}$ at nearly zero temperature and in on-resonance frequency. As is clear, our magnetometer would be two orders of magnitude more sensitive than the case of coplanar microwave waveguide. Comparing the magnon mode and cavity decay rates, which has the same order in these two systems, one can conclude that the main reason of high difference between cooperativities and also magnetic field sensitivity in these two systems is the total spin number. The total spin number in the microwave cavity is about three orders of magnitude larger than the coplanar waveguides, which leads to a higher cooperativity and hence more sensitive magnetometer. Therefore, we emphasize that precise magnetic field measurement requires a small magnon mode decay rate and large total spin number, which both lead to larger cooperativity. Nevertheless, as we know, although a large factor of   $\frac{\sqrt{\kappa_m}}{\eta}$ and hence high cooperativity or significant cavity enhancement has not been achieved yet for magnon-photon interaction based on coplanar microwave waveguides, but finding or fabricating a ferromagnetic material with high spin density and low magnon mode dissipation rate for overcoming these limitations and accessing on-chip integrated precise magnetometer can be promising.

Let us now compare our proposed magnetometer with other ones, especially current state-of-the-art commercial magnetometers such as SQUIDs and atomic-based magnetometers. The SQUIDs and atomic-based magnetometers have sensitivities in the order of  $(0.1$--$1)~ {\rm fT}/\sqrt{\rm Hz}$ \cite{12,13,14,five}. Although they are so sensitive, there are two important challenges with them \cite{13,15}: 
SQUID-based magnetometers cannot work sensitively at room temperature. In fact, they need  cryogenic systems in order to be cooled because as the temperature increases their sensitivity decreases \cite{15}. On the other hand, the sensitivity of the atomic-based magnetometers decreases at lower frequencies \cite{13}. This means that they can work well at high frequencies and, thus, they cannot work over a wide range of frequencies.
Besides, it has been shown \cite{10} that the quantum optomechanical systems with no cooling or cryogenic system can be engineered to operate as a sensitive magnetometer in a wide range of frequencies $2~\rm{Hz}$-$1~\rm{kHz}$ with sensitivity in the order of $150~{\rm nT}/\sqrt{\rm Hz}$ \cite{10}, which is appropriate for some significant applications including magnetic anomaly detection \cite{40}, geophysical surveys \cite{41}, and magnetoencephalography \cite{42}. Also, the sensitivities of the order of $200$, $131 $ and $26~ {\rm pT}/\sqrt{\rm Hz}$  respectively at frequencies about $17\rm MHz$, $127\rm kHz$ and $\rm 11MHz$  were achievable in optomechanic-based magnetometers \cite{10,four, highestsensitive optomechanics}.
Although the proposed optomechanical-based magnetometers can beat the mentioned challenges \cite{10,four}, their sensitivity is still less than the SQUIDs and atomic-based magnetometers.
In addition to the cavity optomechanical systems, recently, some precise magnetometers based on cavity magnon--polariton have been proposed \cite{one,8}. It has been shown that they have the capability to achieve  sensitivity on the order of $1~{\rm fT}/\sqrt{\rm Hz}$ at room temperature \cite{8}. Moreover, the sensitivity in the order of $2.4 ~{\rm pT}/\sqrt{\rm Hz}$ in some magnetometers based on cavity magnon--polariton has been estimated \cite{one}.
Beside these systems, phase estimation algorithm-based magnetometer with sensitivity of the order of $20.7~ \rm {pT}/\sqrt{\rm Hz}$ has also been reported in Ref.~\cite{magnetometry Sorin}.


It is  also worth  comparing the sizes of different magnetometers. Our proposed cavity electromagnonic magnetometer, with its requirement of high spin density, has a large size, especially compared to other magnetometers such as SQUIDs and cavity optomechanics, which have sizes in 1D  respectively on the order of $1$ \cite{size of SQUID} and $10~\rm{\mu m}$ \cite{10} . Although our proposed magnetometer has a larger size in 1D, on the order of  $1$--$10~ mm$,  it has some advantages such as higher sensitivity (on the order of $5.90\times 10^{-18}\frac{\rm T}{\mathrm{\sqrt{Hz}}}$) than the other devices and its ability to work at room temperature and in a wider frequency band (up to  $\rm{MHz}$). These can  motivate their use in some practical applications. In other words, there is a trade-off between the higher sensitivity with the larger size in this scheme and the integrated size with smaller sensitivities in other devices, i.e., \textit{higher sensitivity at the cost of larger size}. It is worth emphasizing that to avoid and decrease the large size in our system, a  ferromagnetic material with higher spin density and lower magnon decay rate should be found or fabricated, which is a challenging issue. Although our proposed magnetometer has a relatively large size with a volume of about $112~ \rm {mm^3}$ and sensitivity of the order of $5.90\times 10^{-18}\frac{\rm T}{\mathrm{\sqrt{Hz}}}$,  based on the reported results of Ref.~ \cite{13}, it is still  around four times smaller and 27 times more sensitive than the current commercial atomic-based magnetometers, such as  one of the most popular precise magnetometers, with volume and sensitivity respectively on the order of $0.45 ~\rm {cm^3}$and  $160 \times 10^{-18} \frac{\rm T}{\sqrt{\rm Hz}}$.

Surprisingly, although it has a relatively large size, according to the theoretical estimated sensitivities, our electromagnetic-based magnetic \textit{amplifier}  sensor, notably beyond the RWA, can operate as a very high-sensitive magnetometer with high sensitivity in the order of $ 10^{-18}~ \rm{T}/\sqrt{\rm Hz} $ at zero temperature which is two or three orders of magnitude better than the current SQUIDs and atomic-based magnetometers. Note that at room temperature our sensor still operates with high sensitivity of the order of $ 0.1 ~ \mathrm{fT /\sqrt{\rm Hz}} $ which is in the same order of the SQUIDs and atomic-based magnetometers.
More interestingly, our proposed magnetic sensor can surpass the current challenges to the SQUIDs and atomic systems because it can operate with very high sensitivity at room temperature and also can work in a very wide range of frequencies $\omega \le\ \kappa_m \sim 20 \rm MHz$ without cooling compared to the optomechanics, SQUIDs, and atomic systems. This proposed magnetometer is even more sensitive than the recent proposed ones based on cavity magnon--polariton and also can work in a wider frequency range.
In fact, by considering the available facilities, there is a trade-off between the higher sensitivity in a wider frequency range with the larger size in the present scheme and the miniaturized sized with smaller sensitivities in other magnetic field sensors.
Therefore, these features of the presented magnetic sensor open another platform for a commercial magnetometer as well as a device for researching the fundamental physics issues like dark matter axions \cite{one, two, three} in the future.


\section{\label{sec6} SUMMARY, CONCLUSION, and outlooks}

In the present contribution, a scheme of a magnetic sensor based on the cavity electromagnonic has been proposed in which magnons and photons interact via the magnetic dipole interaction. It has been shown that beyond the RWA by choosing the suitable system parameters, one can simultaneously achieve  signal amplification and added noise suppression below the SQL in order to precise magnetic measurement. The advantages of the proposed sensor is that our magnetometer can work in a wide range of frequencies at the room temperature with high sensitivity, which is even better than the current SQUIDs  and atomic-based magnetometers that need cooling and work at high-frequency regimes, respectively. 
These properties means that the proposed sensor can be competitive compared to the SQUIDs and atomic-based systems and might be commercial in the future, or used for fundamental applications such as searching the dark matter axion.

Instead of magnetic sensing using the output phase quadrature of cavity mode, the \textit{single-quadrature }magnetic sensing, one can use the generalized rotated output cavity quadrature to exploit the \textit{ponderomotive} squeezing in order to enhance noise reduction and signal response amplification. Furthermore, considering the squeezed vacuum injection instead of thermal noise for the microwave mode may lead to optimization and more controllability in order to achieve more noise suppression and signal amplification. Moreover, in real applications, one needs to consider the classical laser phase noises as presented in Ref.~\cite{Review of force sensing}. Finally, the physical interpretation of the proposed scheme can be deeply analyzed using the introduced  approach of  Green's function in Ref.~\cite{aliGreen}.

\section{AUTHOR CONTRIBUTIONS}
AMF proposed and developed the primary idea of quantum magnetic-sensing and MSE organized the idea in the cavity quantum electromagnonics. All calculations have been performed and checked by MSE and AMF, respectively. All authors contributed to prepare the manuscript. MSE and AMF had equal contributions to write the manuscript. MSE and AMF revised and answered equally to the referees' comments during the referee process. All numerical calculations and graphics have been done by MSE. MBH is group supervisor and supervisor of MSE’s PhD thesis, and also AMF is advisor of MSE’s thesis.

\section{acknowledgments}
The authors thank Prof. Yasunobu Nakamura and also Prof. David Vitali for reading the manuscript and providing helpful suggestions. They would also like to express their gratitude to the referees whose valuable comments have substantially improved the article.
A.M.F. wishes to thank the Office of Graduate Studies of the University of Isfahan (OGSUI) and ICQTs for their support.

\section*{\label{sec7} APPENDIX: MAGNETOMETRY  under the RWA}

Under the RWA, the Hamiltonian of the system is given by 
\begin{eqnarray}\label{56}
\hat{H}&=& \hbar \omega_a\hat{a}^{\dagger}\hat{a}+\hbar \omega_m\hat{m}^{\dagger}\hat{m}+\hbar g_0 (\hat{a}^{\dagger}\hat{m}+\hat{a}\hat{m}^{\dagger})\nonumber\\
&&+i\hbar \epsilon_L(\hat{a}^{\dagger}e^{-i\omega_Lt}-\hat{a}e^{i\omega_L t})+i\hbar \epsilon_d(\hat{m}^{\dagger}e^{-i\omega_d t}-\hat{m}e^{i\omega_d t})\nonumber\\
&&-\hbar \eta B(t) (\hat{m}+\hat{m}^{\dagger}).
\end{eqnarray}
By rewriting the above Hamiltonian in the frame rotating at frequencies $\omega_L$ and $\omega_d$, we get 
\begin{eqnarray}\label{57}
\hat{H}^{(\mathrm{rot})}&=&\hbar\Delta(\hat{a}^{\dagger}\hat{a}+\hat{m}^{\dagger}\hat{m})+\hbar g_0 (\hat{a}^{\dagger}\hat{m}+\hat{a}\hat{m}^{\dagger})+i\hbar\epsilon_L(\hat{a}^{\dagger}-\hat{a})\nonumber\\
&&+i\hbar \epsilon_d(\hat{m}^{\dagger}-\hat{m})-\hbar \eta B(t) (\hat{m}e^{-i\omega_d t}+\hat{m}^{\dagger}e^{i\omega_d t}),\
\end{eqnarray}
in which we have assumed $\omega_L=\omega_d$, and since $\omega_a\simeq\omega_m$, thus $\Delta_a\simeq \Delta_m\equiv \Delta$. Following the method used for the system beyond the RWA, one can obtain the output phase quadrature under the RWA in the Fourier space as follows:
\begin{eqnarray}\label{58}
\hat{P}^{\mathrm{out}}_a(\omega)&=& \mathcal{A}^{\prime}(\omega)\hat{X}^{\prime \mathrm{in}}_m(\omega)+\mathcal{B}^{\prime}(\omega)\hat{P}^{\prime \mathrm{in}}_m(\omega)\nonumber\\
&&+\mathcal{C}^{\prime}(\omega)\hat{X}^{\mathrm{in}}_a(\omega)+\mathcal{D}^{\prime}(\omega)\hat{P}^{\mathrm{in}}_a(\omega)
\end{eqnarray}
where $\hat{X}^{\prime \mathrm{in}}_m(\omega)$ and  $\hat{P}^{\prime \mathrm{in}}_m(\omega)$ are defined as in Eqs.~(\ref{19}) and (\ref{20}) and $\mathcal{A}^{\prime}(\omega)$, $\mathcal{B}^{\prime}(\omega)$, $\mathcal{C}^{\prime}(\omega)$, and $\mathcal{D}^{\prime}(\omega)$ are defined as 
\begin{widetext}
\begin{eqnarray}
 \mathcal{A}^{\prime}(\omega)&  = &  \psi (\omega)\Bigg\{ \frac{g_0\Delta \chi_m(\omega)\chi_a(\omega)\sqrt{\kappa_a\kappa_m}}{(i\omega+ \frac{\kappa_a}{2})}\Bigg[1+\chi_m(\omega)\Bigg(i\omega+\frac{\kappa_a}{2}\Bigg)\Bigg]-\frac{g_0\chi_m(\omega)\sqrt{\kappa_a\kappa_m}}{(i\omega+\frac{\kappa_a}{2})}\Bigg\}, \label{59}\\
\mathcal{B}^{\prime} (\omega)& =&\psi(\omega)\Bigg\{ \frac{g_0\Delta \chi_a(\omega)\sqrt{\kappa_a}}{(i\omega+\frac{\kappa_a}{2})}\Bigg[1+\chi_m(\omega)\Bigg(i\omega+\frac{\kappa_a}{2}\Bigg)\Bigg]\Bigg[-g_0^2\chi_m^3(\omega)\chi_a(\omega)\sqrt{\kappa_m}\Bigg(i\omega+\frac{\kappa_a}{2}\Bigg)+\Delta \chi_m^2(\omega)\sqrt{\kappa_m}\Bigg]
-g_0\chi_m(\omega)\chi_a(\omega)\sqrt{\kappa_a\kappa_m}\nonumber\\
&&(1+\Delta^2\chi_m^2(\omega)-\Delta g_0^2\chi_m^2(\omega)\chi_a (\omega)\Bigg[1
+\chi_m(\omega)\Bigg(i\omega+\frac{\kappa_a}{2}\Bigg)\Bigg]-\frac{g_0\sqrt{\kappa_a}}{(i\omega+\frac{\kappa_a}{2})}
\Bigg[-g_0^2\chi_m^3(\omega)\chi_a(\omega)\sqrt{\kappa_m}\Bigg(i\omega+\frac{\kappa_a}{2}\Bigg)+\Delta\chi_m^2(\omega)\sqrt{\kappa_m}\Bigg]\Bigg\}, \label{60}\\
\mathcal{C}^{\prime}(\omega)&=&\psi(\omega)\Bigg [-g_0^2\chi_m^2(\omega)\chi_a^2(\omega)\Delta \kappa_a \Bigg[1+\chi_m(\omega)\Bigg(i\omega+\frac{\kappa_a}{2}\Bigg)\Bigg]
-\kappa_a\chi_a(\omega)
\Bigg(1+\Delta^2\chi_m^2(\omega)-\Delta g_0^2\chi_m^2(\omega)\chi_a(\omega)\Bigg[1
+\chi_m(\omega)\Bigg(i\omega+\frac{\kappa_a}{2}\Bigg)\Bigg]\Bigg) \nonumber\\    
&&+g_0^2\chi_m^2(\omega)\chi_a(\omega)\kappa_a \Bigg], \label{61}\\
\mathcal{D}^{\prime}(\omega)&=&\psi(\omega)\Bigg\{\frac{-g_0^2\Delta^2\chi_m^2(\omega)\chi_a^2(\omega)\kappa_a}{(i\omega+\frac{\kappa_a}{2})}\Bigg[1+\chi_m(\omega)\Bigg(i\omega+\frac{\kappa}{2}\Bigg)\Bigg]
+\frac{\kappa_a-\Delta\chi_a(\omega)\kappa_a}{(i\omega+\frac{\kappa_a}{2})}
\Bigg[1+\Delta^2\chi_m^2(\omega)-\Delta g_0^2\chi_m^2(\omega)\chi_a(\omega)\Bigg(1
+\chi_m(\omega)\Bigg(i\omega+\frac{\kappa_a}{2}\Bigg) \Bigg)\Bigg] \nonumber\\
&&+\frac{\Delta \kappa_a g_0^2\chi_m^2(\omega)\chi_a(\omega)}{(i\omega+\frac{\kappa_a}{2})}\Bigg\}-1, \label{62}
\end{eqnarray}
\end{widetext}
where we have defined $\psi(\omega)$, $\chi_a(\omega)$, and $\chi_m(\omega)$ as follows
\begin{eqnarray}
\psi(\omega)&=& \Bigg\{1+\Delta^2\chi_m^2(\omega)-\Delta g_0^2\chi_m^2(\omega)\chi_a(\omega)\Bigg[1+\chi_m(\omega)\Bigg(i\omega+\frac{\kappa_a}{2}\Bigg) \Bigg]\nonumber\\
&&-\frac{\Delta g_0^2 \chi_m(\omega)\chi_a(\omega)}{(i\omega+\frac{\kappa_a}{2})}
\Bigg[1+\chi_m(\omega)\Bigg(i\omega+\frac{\kappa_a}{2}\Bigg) \Bigg]+\frac{g_0^2\chi_m(\omega)}{(i\omega+\frac{\kappa_a}{2})}\Bigg\}^{-1}, \nonumber\\\label{63} \\
\chi_a(\omega)&=&\Delta \Bigg[\Bigg(i\omega+\frac{\kappa_a}{2}\Bigg)^2+\Delta^2+g_0^2\chi_m(\omega)\Bigg(i\omega+\frac{\kappa_a}{2}\Bigg)\Bigg]^{-1},\label{64}\\
\chi_m(\omega)&=&\Bigg[i\omega+\frac{\kappa_m}{2}\Bigg]^{-1}.\label{65}
\end{eqnarray}

Therefore, analogous to Eq.~(\ref{29}), the spectrum of the optical output phase quadrature can be obtained as follows:
\begin{eqnarray}\label{66}
S_{P^{\mathrm{out}}_a}(\omega)&=&[\vert \mathcal{A}^{\prime}(\omega)\vert ^2+\vert \mathcal{B}^{\prime}(\omega)\vert ^2](\bar{n}_m+\frac{1}{2})\nonumber\\
&&+ [\vert \mathcal{C}^{\prime}(\omega)\vert ^2+\vert \mathcal{D}^{\prime}(\omega)\vert ^2](\bar{n}_a+\frac{1}{2}),
\end{eqnarray}
which can be rewritten as
\begin{equation}\label{67}
S_{P^{out}_a}(\omega)=R_m^{\prime}(\omega)[(\bar{n}_m+\frac{1}{2})+n_{add}^{\prime}(\omega)],
\end{equation}
where 
\begin{eqnarray}\label{68}
R_m^{\prime}(\omega)=[\vert \mathcal{A}^{\prime}(\omega)\vert ^2+\vert \mathcal{B}^{\prime}(\omega)\vert ^2]
 \end{eqnarray}
and
\begin{eqnarray}\label{69}
 n_{add}^{\prime}(\omega)=(\bar{n}_a+\frac{1}{2})\frac{ [\vert \mathcal{C}^{\prime}(\omega)\vert ^2+\vert \mathcal{D}^{\prime}(\omega)\vert ^2]}{[\vert \mathcal{A}^{\prime}(\omega)\vert ^2+\vert \mathcal{B}^{\prime}(\omega)\vert ^2]}
\end{eqnarray}
are the magnonic response and the added noise, respectively. By choosing the zero detuning for external drive fields and considering the resonance frequency, one can obtain the magnonic response and the added noise as given in the main text in Eqs.~(\ref{37}) and (\ref{38}).

%





\begin{thebibliography}{99} 






\bibitem{1} 
M. Tsang and C. M. Caves,  ``Coherent quantum-noise cancellation for optomechanical sensors,"\href{https://doi.org/10.1103/PhysRevLett.105.123601} {Phys. Rev. Lett. \textbf{105}, 123601 (2010)}.

\bibitem{virgo2020}
F. Acernese, M. Agathos, L. Aiello, A. Ain, A. Allocca, A. Amato, S. Ansoldi, S. Antier, M. Ar`ene, N. Arnaud, et al., ``Quantum Backaction on kg-Scale Mirrors: Observation of Radiation Pressure Noise in the Advanced Virgo Detector," \href{https://doi.org/10.1103/PhysRevLett.125.131101}{Phys. Rev. Lett. \textbf{125}, 131101 (2020)}.


\bibitem{2} 
 X. Xu, and J. M. Taylor, ``Squeezing in a coupled two-mode optomechanical system for force sensing below the standard quantum limit,"\href{https://doi.org/10.1103/PhysRevA.90.043848}{ Phys. Rev. A \textbf{90}, 043848 (2014)}.


\bibitem{3}
  B. Levitan, A. Metelmann, and A. Clerk, ``Optomechanics with two-phonon driving,"\href{https://doi.org/10.1088/1367-2630/18/9/093014} {New J. Phys. \textbf{18}, 093014 (2016)}.

\bibitem{4}  
A. Motazedifard, F. Bemani, M. Naderi, R. Roknizadeh, and D. Vitali, ``Force sensing based on coherent quantum noise cancellation in a hybrid optomechanical cavity with squeezed-vacuum injection,"\href{https://doi.org/10.1088/1367-2630/18/7/073040} {New J. Phys. \textbf{18}, 073040 (2016)}.

\bibitem{5}
  A. Motazedifard, A. Dalafi, F. Bemani, and M. Naderi, ``Force sensing in hybrid Bose-Einstein-condensate optomechanics based on parametric amplification,"\href{https://doi.org/10.1103/PhysRevA.100.023815} {Phy. Rev. A \textbf{100}, 023815 (2019)}.

\bibitem{Review of force sensing}
A. Motazedifard, A. Dalafi, and M. Naderi, ``Ultraprecision quantum sensing and measurement based on nonlinear hybrid optomechanical systems containing ultracold atoms or atomic Bose–Einstein condensate," \href{https://doi.org/10.1116/5.0035952}{AVS Quantum Science \textbf{3}, 024701 (2021)}.

\bibitem{6} 
 J. Lenz and S. Edelstein, ``Magnetic sensors and their applications,"\href{https://doi.org/10.1109/JSEN.2006.874493} {IEEE Sensors journal \textbf{6}, 631-649 (2006)}.

\bibitem{7} 
 D. Budker and M. Romalis, ``Optical magnetometry,"\href{https://doi.org/10.1038/nphys566} {Nat. Phys. \textbf{3}, 227-234 (2007)}.

\bibitem{AVS magnetometry} 
 K.-M. C. Fu, G. Z. Iwata, A. Wickenbrock, and D. Budker, ``Sensitive magnetometry in challenging environments,"\href{ https://doi.org/10.1116/5.0025186}{ AVS Quantum Sci \textbf{2}, 044702 (2020)}.

\bibitem{one}
N. Crescini, C. Braggio, G. Carugno, A. Ortolan, and G. Ruoso, ``Cavity magnon polariton based precision magnetometry ,"\href{ https://doi.org/10.1063/5.0024369} {Applied Physics Letters \textbf{117}, 144001 (2020)}.


\bibitem{two}
P.-H. Chu, Y. J. Kim, and I. Savukov, ``Search for an axion-induced oscillating electric dipole moment for electrons using atomic magnetometers,"\href{https://doi.org/10.1103/PhysRevD.99.075031}{Phys. Rev D \textbf{99}, 075031 (2019)}.

\bibitem{three}
A. V. Gramolin, D. Aybas, D. Johnson, J. Adam, and A. O. Sushkov,``Search for axion-like dark matter with ferromagnets," \href{https://doi.org/10.1038/s41567-020-1006-6}{Nat. Phys. \textbf{17}, 79 (2021)}.



\bibitem{phasemodulatedmagnetometry}
N. Crescini, G. Carugno, and G. Ruoso, ``Phase-modulated cavity magnon polaritons as a precise magnetic field probe,"\href{https://arxiv.org/abs/2010.00093v2}{ arXiv:2010.00093v2}.

\bibitem{8} 
 Y. Cao, and P. Yan, ``Exceptional magnetic sensitivity of PT-symmetric cavity magnon polaritons,"\href{https://doi.org/10.1103/PhysRevB.99.214415} {Phys. Rev. B \textbf{99}, 214415 (2019)}.

\bibitem{9} 
S. Forstner, S. Prams, J. Knittel, E. Van Ooijen, J. Swaim, G. Harris, A. Szorkovszky, W. Bowen, and H. Rubinsztein-Dunlop,``Cavity optomechanical magnetometer," \href{https://doi.org/10.1103/PhysRevLett.108.120801}{Phys. Rev. Lett. \textbf{108}, 120801 (2012)}.

\bibitem{10}
 S. Forstner, E. Sheridan, J. Knittel, C. L. Humphreys, G. A. Brawley, H. Rubinsztein‐Dunlop, and W. P. Bowen, ``Ultrasensitive optomechanical magnetometry,"\href{  https://doi.org/10.1002/adma.201401144} {Adv. Mater. \textbf{26}, 6348-6353 (2014)}.

\bibitem{11} 
 B.-B. Li, J. Bilek, U. B. Hoff, L. S. Madsen, S. Forstner, V. Prakash, C. Schäfermeier, T. Gehring, W. P. Bowen, and U. L. Andersen, ``Quantum enhanced optomechanical magnetometry,"\href{https://doi.org/10.1364/OPTICA.5.000850} {Optica \textbf{5}, 850-856 (2018)}.

\bibitem{four} 
C. Yu, J. Janousek, E. Sheridan, D. L. McAuslan, H. Rubinsztein-Dunlop, P. K. Lam, Y. Zhang, and W. P. Bowen, ``Optomechanical magnetometry with a macroscopic resonator," \href{https://doi.org/10.1103/PhysRevApplied.5.044007}{Phys. Rev. Applied \textbf{5}, 044007 (2016)}.


\bibitem{12} 
 I. Kominis, T. Kornack, J. Allred, and M. V. Romalis, ``A subfemtotesla multichannel atomic magnetometer,"\href{https://doi.org/10.1038/nature01484} {Nature (London) \textbf{422}, 596-599 (2003)}.

\bibitem{13}
 H. Dang, A. C. Maloof, and M. V. Romalis, ``Ultrahigh sensitivity magnetic field and magnetization measurements with an atomic magnetometer,"\href{https://doi.org/10.1063/1.3491215} {Appl. Phys. Lett. \textbf{97}, 151110 (2010)}.


\bibitem{Noisy atomic magnetometry in real time}
J. A. Binefa and J. Kolodynski, ``Noisy atomic magnetometry in real time," \href{https://arxiv.org/abs/2103.12025} {arXiv preprint arXiv:2103.12025  (2021)}.


\bibitem{14}
 R. Kleiner, D. Koelle, F. Ludwig, and J. Clarke, ``Superconducting quantum interference devices: State of the art and applications,"\href{https://doi.org/10.1109/JPROC.2004.833655} {Proceedings of the IEEE 92, 1534 (2004)}.

\bibitem{15} 
 J. Gallop, ``SQUIDs: Some limits to measurement,"\href{https://doi.org/10.1088/0953-2048/16/12/055} {Supercond. Sci. Technol. \textbf{16}, 1575 (2003)}.

\bibitem{size of SQUID} 
F. Baudenbacher, L. Fong, J. Holzer, and M. Radparvar, ``Monolithic low-transition-temperature superconducting magnetometers for high resolution imaging magnetic fields of room temperature samples ,"\href{ https://doi.org/10.1063/1.1572968}{Appl. Phys. Lett. \textbf{82}, 3487 (2003)}.


\bibitem{16} 
 Y. Tabuchi, S. Ishino, T. Ishikawa, R. Yamazaki, K. Usami, and Y. Nakamura, ``Hybridizing ferromagnetic magnons and microwave photons in the quantum limit,"\href{https://doi.org/10.1103/PhysRevLett.113.083603} {Phys. Rev. Lett. \textbf{113}, 083603 (2014)}.

\bibitem{17} 
 X. Zhang, C.-L. Zou, L. Jiang, and H. X. Tang, ``Strongly coupled magnons and cavity microwave photons,"\href{https://doi.org/10.1103/PhysRevLett.113.156401} {Phys. Rev. Lett. \textbf{113}, 156401 (2014)}. note that the experimental parameters for obtaining the sensitivity under the RWA are choosen from the  version on arxiv (https://arxiv.org/abs/1405.7062).

\bibitem{18} 
M. Goryachev, W. G. Farr, D. L. Creedon, Y. Fan, M. Kostylev, and M. E. Tobar, ``High-cooperativity cavity QED with magnons at microwave frequencies,"\href{https://doi.org/10.1103/PhysRevApplied.2.054002}{Phys. Rev. Applied. \textbf{2}, 054002 (2014)}.

\bibitem{19} 
T. Liu, X. Zhang, H. X. Tang, and M. E. Flatté, ``Optomagnonics in magnetic solids,"\href{https://doi.org/10.1103/PhysRevB.94.060405} {Phys. Rev. B \textbf{94}, 060405(R) (2016)}.

\bibitem{20} 
 S. V. Kusminskiy, H. X. Tang, and F. Marquardt, ``Coupled spin-light dynamics in cavity optomagnonics,"\href{https://doi.org/10.1103/PhysRevA.94.033821} {Phys. Rev. A \textbf{94}, 033821 (2016)}.

\bibitem{21}
  S. Sharma, Y. M. Blanter, and G. E. Bauer, ``Light scattering by magnons in whispering gallery mode cavities,"\href{https://doi.org/10.1103/PhysRevB.96.094412} {Phys. Rev. B \textbf{96}, 094412 (2017)}.

\bibitem{22} 
 X. Zhang, N. Zhu, C.-L. Zou, and H. X. Tang, ``Optomagnonic whispering gallery microresonators,"\href{https://doi.org/10.1103/PhysRevLett.117.123605} {Phys. Rev. lett. \textbf{117}, 123605 (2016)}.

\bibitem{23} 
 A. Osada, R. Hisatomi, A. Noguchi, Y. Tabuchi, R. Yamazaki, K. Usami, M. Sadgrove, R. Yalla, M. Nomura, and Y. Nakamura, ``Cavity optomagnonics with spin-orbit coupled photons,"\href{https://doi.org/10.1103/PhysRevLett.116.223601} {Phys. Rev. lett. \textbf{116}, 223601 (2016)}.

\bibitem{24} 
 A. Serga, A. Chumak, and B. Hillebrands, ``YIG magnonics,"\href{https://doi.org/10.1088/0022-3727/43/26/264002} { J. Phys. D: Appl. Phys. \textbf{43}, 264002 (2010)}.

\bibitem{25}
 A. V. Chumak, V. I. Vasyuchka, A. A. Serga, and B. Hillebrands, ``Magnon spintronics,"\href{https://doi.org/10.1038/nphys3347} {Nat. Phys. \textbf{11}, 453-461 (2015)}.


\bibitem{26}
D. D. Stancil, and A. Prabhakar, \textit{Spin waves} (Springer, Berlin, 2009).

\bibitem{magnon sensing} 
S. P. Wolski, D. Lachance-Quirion, Y. Tabuchi, S. Kono, A. Noguchi, K. Usami, and Y. Nakamura, ``Dissipation-based quantum sensing of magnons with a superconducting qubit,"\href{https://doi.org/10.1103/PhysRevLett.125.117701}{Phys. Rev. Lett.\textbf{125}.117701  (2020)}.

  
\bibitem{Nori} 
D. Zhang, X.-M. Wang, T.-F. Li, X.-Q. Luo, W. Wu, F. Nori, and J. You,``Cavity quantum electrodynamics with ferromagnetic magnons in a small yttrium-iron-garnet sphere," \href{https://doi.org/10.1038/npjqi.2015.14}{npj Quantum Information \textbf{1}, 1 (2015)}.
  
\bibitem{Highcooperativity}
H. Huebl, C. W. Zollitsch, J. Lotze, F. Hocke, M. Greifenstein, A. Marx, R. Gross, and S. T. Goennenwein,``High cooperativity in coupled microwave resonator ferrimagnetic insulator hybrids," \href{https://doi.org/10.1103/PhysRevLett.111.127003}{Phys. Rev. Lett. \textbf{111}, 127003 (2013)}.
  

\bibitem{28}
 D. Lachance-Quirion, Y. Tabuchi, A. Gloppe, K. Usami, and Y. Nakamura, ``Hybrid quantum systems based on magnonics,"\href{https://doi.org/10.7567/1882-0786/ab248d} {Appl. Phys. Expr \textbf{12}, 070101 (2019)}.

\bibitem{29}
 X. Zhang, C.-L. Zou, L. Jiang, and H. X. Tang, ``Cavity magnomechanics,"\href{https://doi.org/10.1126/sciadv.1501286 }{Science advances \textbf{2}, e1501286 (2016)}.


\bibitem{Dynamical Backaction Magnomechanics}
C. Potts, E. Varga, V. Bittencourt, S. V. Kusminskiy, and J. Davis,``Dynamical backaction magnomechanics," \href{https://arxiv.org/abs/2104.11218} {arXiv preprint arXiv:2104.11218  (2021)}.


\bibitem{30}
R. Hisatomi, A. Osada, Y. Tabuchi, T. Ishikawa, A. Noguchi, R. Yamazaki, K. Usami, and Y. Nakamura, ``Bidirectional conversion between microwave and light via ferromagnetic magnons"\href{https://doi.org/10.1103/PhysRevB.93.174427}{Phys. Rev. B \textbf{93}, 174427 (2016)}.


\bibitem{31}
  J. Li, S.-Y. Zhu, and G. Agarwal, ``Magnon-photon-phonon entanglement in cavity magnomechanics,"\href{https://doi.org/10.1103/PhysRevLett.121.203601} {Phys. Rev. Lett. \textbf{121}, 203601 (2018)}.

\bibitem{32} 
 M. Yu, H. Shen, and J. Li, ``Magnetostrictively induced stationary entanglement between two microwave fields,"\href{https://doi.org/10.1103/PhysRevLett.124.213604} {Phys. Rev. Lett. \textbf{124}, 213604 (2020)}.

\bibitem{33}
 M. Yu, S.-Y. Zhu, and J. Li, ``Macroscopic entanglement of two magnon modes via quantum correlated microwave fields,"\href{https://doi.org/10.1088/1361-6455/ab68b5} { J. Phys. B: At. Mol. Opt. Phys. \textbf{53}, 065402 (2020)}.

\bibitem{34}
J. Li and S. Groeblacher, ``Entangling the vibrational modes of two massive ferromagnetic spheres using cavity magnomechanics,"\href{https://doi.org/10.1088/2058-9565/abd982}{Quantum Sci. Technol. \textbf{6}, 024005   (2021)}.

\bibitem{35}
J. Li and S.-Y. Zhu,``Entangling two magnon modes via magnetostrictive interaction"\href{https://doi.org/10.1088/1367-2630/ab3508}{New J. Phys. \textbf{21}, 085001 (2019)}.

\bibitem{36} 
 J. M. P. Nair, and G. S. Agarwal, ``Quantum drives produce strong entanglement between YIG samples without using intrinsic nonlinearities,"\href{https://arxiv.org/abs/1905.07884} {arXiv preprint arXiv:1905.07884 (2019)}.


\bibitem{last} 
Z. Zhang, M. O. Scully, and G. S. Agarwal, ``Quantum entanglement between two magnon modes via Kerr nonlinearity driven far from equilibrium,"\href{https://doi.org/10.1103/PhysRevResearch.1.023021}{Phys. Rev. Research \textbf{1}, 023021 (2019)}.


\bibitem{electromagnonics-optomechanics} 
H. Tan and J. Li,```Einstein-Podolsky-Rosen entanglement and asymmetric steering between distant macroscopic mechanical and magnonic systems,"\href{https://doi.org/10.1103/PhysRevResearch.3.013192} {Phys. Rev. Research \textbf{3}, 013192 (2021)}.

\bibitem{foroudcrystalentanglement} F. Bemani, R. Roknizadeh, A. Motazedifard, M. H. Naderi, and D. Vitali, 
\href{https://doi.org/10.1103/PhysRevA.99.063814}{ Phys. Rev. A \textbf{99}, 063814 (2019)}.


\bibitem{37} 
J. Li, Y.-P. Wang, J. You, and S.-Y. Zhu, ``Squeezing microwave fields via magnetostrictive interaction,"\href{https://arxiv.org/abs/2101.02796}{ arXiv:2101.02796}.


\bibitem{microwace field squeezing} 
 J. Li, S.-Y. Zhu, and G. Agarwal, ``Squeezed states of magnons and phonons in cavity magnomechanics,"\href{https://doi.org/10.1103/PhysRevA.99.021801} {Phys. Rev. A \textbf{99}, 02180(R) (2019)}.
 

\bibitem{38} 
 M.-S. Ding, L. Zheng, and C. Li, ``Phonon laser in a cavity magnomechanical system,"\href{https://doi.org/10.1038/s41598-019-52050-7} {Sci. Rep. \textbf{9}, 1 (2019)}.


\bibitem{39}
  C. Potts, V. A. Bittencourt, S. V. Kusminskiy, and J. Davis, ``Magnon-phonon quantum correlation thermometry,"\href{https://doi.org/10.1103/PhysRevApplied.13.064001} {Phys. Rev.  Applied \textbf{13}, 064001 (2020)}.


\bibitem{magnonblockade1}
Z.-X. Liu, H. Xiong, and Y. Wu, ``Magnon blockade in a hybrid ferromagnet-superconductor quantum system," \href{https://doi.org/10.1103/PhysRevB.100.134421} {Phys. Rev. B \textbf{100}, 134421 (2019)}.



\bibitem{magnonblockade2}
J.-k. Xie, S.-l. Ma, and F.-l. Li, ``Quantum-interference-enhanced magnon blockade in an yttrium-iron-garnet sphere coupled to superconducting circuits,"\href{https://doi.org/10.1103/PhysRevA.101.042331} {Phys. Rev. A \textbf{101}, 042331 (2020)}.


\bibitem{quantumillumination}
Q. Cai, J. Liao, B. Shen, G. Guo, and Q. Zhou, ``Microwave quantum illumination via cavity magnonics,"\href{https://arxiv.org/abs/2011.04301} {arXiv:2011.04301}.


\bibitem{photon-phononconversion}
S.-f. Qi and J. Jing,  ``Magnon-assisted photon-phonon conversion in the presence of the structured environments,"\href{https://doi.org/10.1103/PhysRevA.103.043704} {Phys. Rev. A \textbf{103}, 043704 (2021)}.


\bibitem{Cavity magnomechanical storage and retrieval of quantum states}
B. Sarma, T. Busch, and J. Twamley,``Cavity magnomechanical storage and retrieval of quantum states,"\href{https://doi.org/10.1088/1367-2630/abf535} {New J. Phys. \textbf{23} 043041  (2021)}.


\bibitem{27}  
Y. Tabuchi, S. Ishino, A. Noguchi, T. Ishikawa, R. Yamazaki, K. Usami, and Y. Nakamura, ``Quantum magnonics: The magnon meets the superconducting qubit,"\href{https://doi.org/10.1016/j.crhy.2016.07.009} {Compt. Ren. Phys. \textbf{17}, 729-739 (2016)}.

\bibitem{USC}
P. Forn-Díaz, L. Lamata, E. Rico, J. Kono, and E. Solano, ``Ultrastrong coupling regimes of light-matter interaction,"\href{https://doi.org/10.1103/RevModPhys.91.025005}{Rev. Mod. Phys 91, 025005 (2019)}.



\bibitem{aliDCEsqueezing} A. Motazedifard, A. Dalafi, M. Naderi, and R. Roknizadeh, ``Strong quadrature squeezing and quantum amplification in a coupled Bose–Einstein condensate—optomechanical cavity based on parametric modulation,"\href{https://doi.org/10.1016/j.aop.2019.03.019}{Annals of Physics \textbf{405}, 202 (2019)}.

\bibitem{aliDCE1} A. Motazedifard, M. H. Naderi, and R. Roknizadeh, ``Analogue model for controllable Casimir radiation in a nonlinear cavity with amplitude-modulated pumping:  Generation and quantum statistical properties," 
\href{https://doi.org/10.1364/JOSAB.32.001555}{J. Opt. Soc. Am. B \textbf{32}, 1555 (2015)}.


\bibitem{aliDCE2} A. Motazedifard, M. H. Naderi, and R. Roknizadeh, ``Dynamical Casimir effect of phonon excitation in the dispersive regime of cavity optomechanics,"
\href{https://doi.org/10.1364/JOSAB.34.000642}{J. Opt. Soc. Am. B \textbf{34}, 642 (2017)}.


\bibitem{aliDCE3} A. Motazedifard, A. Dalafi, M. H. Naderi, and R. Roknizadeh,``Controllable generation of photons and phonons in a coupled Bose-Einstein condensate-optomechanical cavity via the parametric dynamical Casimir effect," \href{https://doi.org/10.1016/j.aop.2018.07.013}{Ann. Phys. \textbf{396}, 202 (2018)}.

\bibitem{NoriDCE1} O. Di Stefano, A. Settineri, V. Macrì, A. Ridolfo, R. Stassi, A. F. Kockum, S. Savasta, and F. Nori, ``Interaction of mechanical oscillators mediated by the exchange of virtual photon pairs," \href{https://doi.org/10.1103/PhysRevLett.122.030402}{Phys. Rev. Lett. \textbf{122}, 030402 (2019)}.

\bibitem{NoriDCE2} W. Qin, V. Macrì, A. Miranowicz, S. Savasta, and F. Nori, ``Experimentally feasible dynamical Casimir effect in parametrically amplified cavity optomechanics," \href{https://arxiv.org/abs/1902.04216v1}{arXiv:1902.04216v1 (2019)}. 

\bibitem{NoriDCE3} 
A. Settineri, V. Macrì, L. Garziano, O. Di Stefano, F. Nori, and S. Savasta, ``Conversion of mechanical noise into correlated photon pairs: Dynamical Casimir effect from an incoherent mechanical drive,"\href{https://doi.org/10.1103/PhysRevA.100.022501} {Phys. Rev. A \textbf{100}, 022501 (2019)}.

\bibitem{pontinmodulation} A. Pontin, M. Bonaldi, A. Borrielli, L. Marconi, F. Marino, G. Pandraud, G. A. Prodi,
P. M. Sarro, E. Serra, and F. Marin, ``Dynamical two-mode squeezing of thermal fluctuations in a cavity optomechanical system," 
\href{https://doi.org/10.1103/PhysRevLett.116.103601}{Phys. Rev. Lett. \textbf{116}, 103601 (2016)}.


\bibitem{ultra strong coupling regime nori}
 A. F. Kockum, A. Miranowicz, S. De Liberato, S. Savasta, and F. Nori, ``Ultrastrong coupling between light and matter,"\href{https://doi.org/10.1038/s42254-018-0006-2}{Nat. Rev. Phys. \textbf{1}, 19 (2019)}.


\bibitem{kalman filtering}
B. Gong, D. Dong, W. Su, and W. Cui, ``Force tracking in cavity optomechanics with a two-level quantum system by Kalman filtering,"\href{https://arxiv.org/abs/1903.01283}{ arXiv:1903.01283  (2019)}.


\bibitem{CPW 2}
Y. Li, T. Polakovic, Y. L. Wang, J. Xu, S. Lendinez, Z. Zhang, J. Ding, T. Khaire, H. Saglam, R.
Divan, J. Pearson, W. K. Kwok, Z. Xiao, V. Novosad, A. Hoffmann, and W. Zhang, ``Strong coupling between magnons and microwave photons in on-chip ferromagnet-superconductor thin-film devices,"\href{https://doi.org/10.1103/PhysRevLett.123.107701}{Phys. Rev. Lett \textbf{123}, 107701 (2019)}.

\bibitem{CPW 3}
J. T. Hou and L. Liu,``Strong coupling between microwave photons and nanomagnet magnons," \href{https://doi.org/10.1103/PhysRevLett.123.107702}{Phys. Rev. Lett \textbf{123}, 107702 (2019)}.


\bibitem{CPW 4}
R. Morris, A. Van Loo, S. Kosen, and A. Karenowska, ``Strong coupling of magnons in a YIG sphere to photons in a planar superconducting resonator in the quantum limit,"\href{https://doi.org/10.1038/s41598-017-11835-4}{Sci. Rep. \textbf{7}, 1 (2017)}.




\bibitem{five}
J. Li, W. Quan, B. Zhou, Z. Wang, J. Lu, Z. Hu, G. Liu, and J. Fang,``SERF Atomic Magnetometer–Recent advances and applications: A review," \href{https://doi.org/10.1109/JSEN.2018.2863707}{IEEE Sensors J.\textbf{18}, 8198 (2018)}.


\bibitem{40}
J. Zhai, Z. Xing, S. Dong, J. Li, and D. Viehland, ``Detection of pico-Tesla magnetic fields using magneto-electric sensors at room temperature,"\href{https://doi.org/10.1063/1.2172706}{Appl. Phys. Lett. \textbf{88}, 062510 (2006)}.

\bibitem{41}
H. G. Meyer, R. Stolz, A. Chwala, and M. Schulz, ``SQUID technology for geophysical exploration,"\href{ https://doi.org/10.1002/pssc.200460832}{phys. Stat. Solidi (C) \textbf{2}, 1504 (2005)}.

\bibitem{42}
H. Xia, A. Ben-Amar Baranga, D. Hoffman, and M. Romalis, ``Magnetoencephalography with an atomic magnetometer," \href{https://doi.org/10.1063/1.2392722}{Appl. Phys. Lett. \textbf{89}, 211104 (2006)}.

\bibitem{highestsensitive optomechanics}
B.-B. Li, G. Brawley, H. Greenall, S. Forstner, E. Sheridan, H. Rubinsztein-Dunlop, and W. P. Bowen,``Ultrabroadband and sensitive cavity optomechanical magnetometry," \href{https://doi.org/10.1364/PRJ.390261}{Photonics Research \textbf{8}, 1064 (2020)}.

\bibitem{magnetometry Sorin}
S. Danilin, A. V. Lebedev, A. Vepsäläinen, G. B. Lesovik, G. Blatter, and G. Paraoanu, ``Quantum-enhanced magnetometry by phase estimation algorithms with a single artificial atom,"\href{https://doi.org/10.1038/s41534-018-0078-y}{npj Quantum Information 4, 1 (2018)}.


\bibitem{aliGreen} Ali Motazedifard, A. Dalafi, M. H. Naderi, 
\href{https://doi.org/10.1088/1751-8121/abf3e9}{J. Phys. A: Math. Theor. \textbf{54}, 215301 (2021)}.




\end{thebibliography}
\end{document}